\documentclass[11pt]{article}
\usepackage[top=1in,bottom=1in,left=1in,right=1in]{geometry} 
\usepackage[onehalfspacing]{setspace}

\usepackage{hyperref, graphicx, caption, subcaption, xcolor, verbatim, titlesec, booktabs, multirow, enumitem, tocloft, natbib, bbm, mathrsfs}
\hypersetup{pdfborder = {0 0 0},colorlinks=true,allcolors=blue}

\usepackage[utf8]{inputenc}
\usepackage[T1]{fontenc}

\usepackage{amsfonts, amssymb, amsmath, amsthm}
\usepackage{graphicx, caption, subcaption, xcolor, verbatim, multirow, enumitem, natbib, bbm, mathrsfs,tabularx,booktabs,listings}
\usepackage{mathtools}
\usepackage[sans]{dsfont}
\usepackage[scr=rsfs,cal=boondox]{mathalpha}
\usepackage{placeins}
\usepackage{threeparttable}
\usepackage[normalem]{ulem}
\theoremstyle{definition}
\newtheorem{remark_tmp}{Remark}

\theoremstyle{definition}
\allowdisplaybreaks[1]
\DeclareMathOperator*{\argmin}{arg\,min}

\DeclareMathOperator{\setbd}{bd}

\DeclareMathOperator{\CI}{CI}

\renewcommand{\P}{\mathbb{P}}
\newcommand{\E}{\mathbb{E}}
\newcommand{\Var}{\mathbb{V}}

\newcommand{\Indicator}{\mathds{1}}
\newcommand{\dif}{\mathop{}\!\mathrm{d}}

\renewcommand{\d}{\mathcal{d}}

\newcommand{\bb}{\mathbf{b}}

\newcommand{\be}{\mathbf{e}}
\newcommand{\bk}{\mathbf{k}}

\newcommand{\br}{\mathbf{r}}

\newcommand{\bu}{\mathbf{u}}

\newcommand{\bx}{\mathbf{x}}
\newcommand{\bX}{\mathbf{X}}

\newcommand{\bZ}{\mathbf{Z}}

\newcommand{\bnu}{\boldsymbol{\nu}}
\newcommand{\bdelta}{\boldsymbol{\delta}}
\newcommand{\bgamma}{\boldsymbol{\gamma}}
\newcommand{\blambda}{\boldsymbol{\lambda}}
\newcommand{\bGamma}{\boldsymbol{\Gamma}}
\newcommand{\bSigma}{\boldsymbol{\Sigma}}

\newcommand{\A}{\mathcal{A}}
\newcommand{\B}{\mathcal{B}}
\newcommand{\I}{\mathcal{I}}

\definecolor{codegray}{gray}{0.95}

\usepackage{listings}
\definecolor{codegreen}{rgb}{0,0.6,0}
\definecolor{codegray}{rgb}{0.5,0.5,0.5}
\definecolor{codepurple}{rgb}{0.58,0,0.82}
\definecolor{backcolour}{rgb}{0.95,0.95,0.95}
 
\lstdefinestyle{Rstyle}{
	language=R,
    backgroundcolor=\color{backcolour},   
    commentstyle=\color{codegreen},
    numberstyle=\tiny\color{codegray},
    stringstyle=\color{codepurple},
    basicstyle=\footnotesize\ttfamily,
    breakatwhitespace=false,         
    breaklines=true,                 
    captionpos=b,                    
    keepspaces=true,                 
    numbers=none,                    
    numbersep=5pt,                  
    showspaces=false,                
    showstringspaces=false,
    showtabs=false,                  
    tabsize=2,
    upquote=true
}
\newcommand{\proglang}{\texttt}
\newcommand{\pkg}{\texttt}
\newcommand{\code}[1]{\texttt{\detokenize{#1}}}

\title{\texttt{rd2d}: Causal Inference in Boundary Discontinuity Designs\bigskip}
\author{Matias D. Cattaneo\thanks{Department of Operations Research and Financial Engineering, Princeton University.} \and
	    Rocio Titiunik\thanks{Department of Politics, Princeton University.} \and
	    Ruiqi (Rae) Yu\thanks{Department of Operations Research and Financial Engineering, Princeton University.} 
	    }

\clearpage

\begin{document}

\maketitle

\begin{abstract}
  Boundary Discontinuity (BD) designs are used in empirical research to learn about causal treatment effects along a continuous assignment boundary defined by a bivariate score. These designs are also known as multi-score regression discontinuity (RD) designs, and include geographic RD designs as a prominent example. This article introduces \pkg{rd2d}, a statistical software package for \proglang{R}, \proglang{Python}, and \proglang{Stata} that implements local polynomial estimation and inference for BD designs using either the bivariate score or a univariate signed distance-to-boundary score. The software covers sharp and fuzzy BD designs, providing automatic bandwidth selection, robust bias-corrected pointwise inference, uniform confidence bands, cluster-robust inference with joint or separate fitting conventions, covariate-adjusted efficiency improvements, mass-point checks, and covariance regularization, among other features. We illustrate the package with an empirical application to Opportunity Zones, where eligibility has a strong first-stage effect on designation but no significant effects on early workplace-job growth.

\end{abstract}


\section{Introduction}\label{sec: Introduction}

Regression Discontinuity (RD) designs are commonly used for treatment effect estimation and causal inference in quantitative sciences \citep[see][and references therein]{Cattaneo-Titiunik_2022_ARE}. In their canonical form, each unit $i\in \{1,2,\cdots,n\}$ is assigned to control ($T_i=0$) or treatment ($T_i=1$) according to the discontinuous rule $T_i = \Indicator(X_i \geq c)$, where $X_i$ denotes a scalar score variable, $c$ denotes a scalar cutoff, and $\Indicator(\cdot)$ is the indicator function. The key idea underlying all RD designs is that units with a score near the cutoff determining treatment assignment are comparable in terms of all pretreatment observables and unobservable characteristics, the only difference being that some units are assigned to control ($X_i < c$) while others are assigned to treatment ($X_i \geq c$). Therefore, in the absence of score manipulation, units with scores near but on different sides of the cutoff can be used as counterfactuals for each other to learn about causal treatment effects.

\begin{figure}
    \centering
    \begin{subfigure}[b]{0.45\textwidth}
        \centering
        \includegraphics[width=\linewidth]{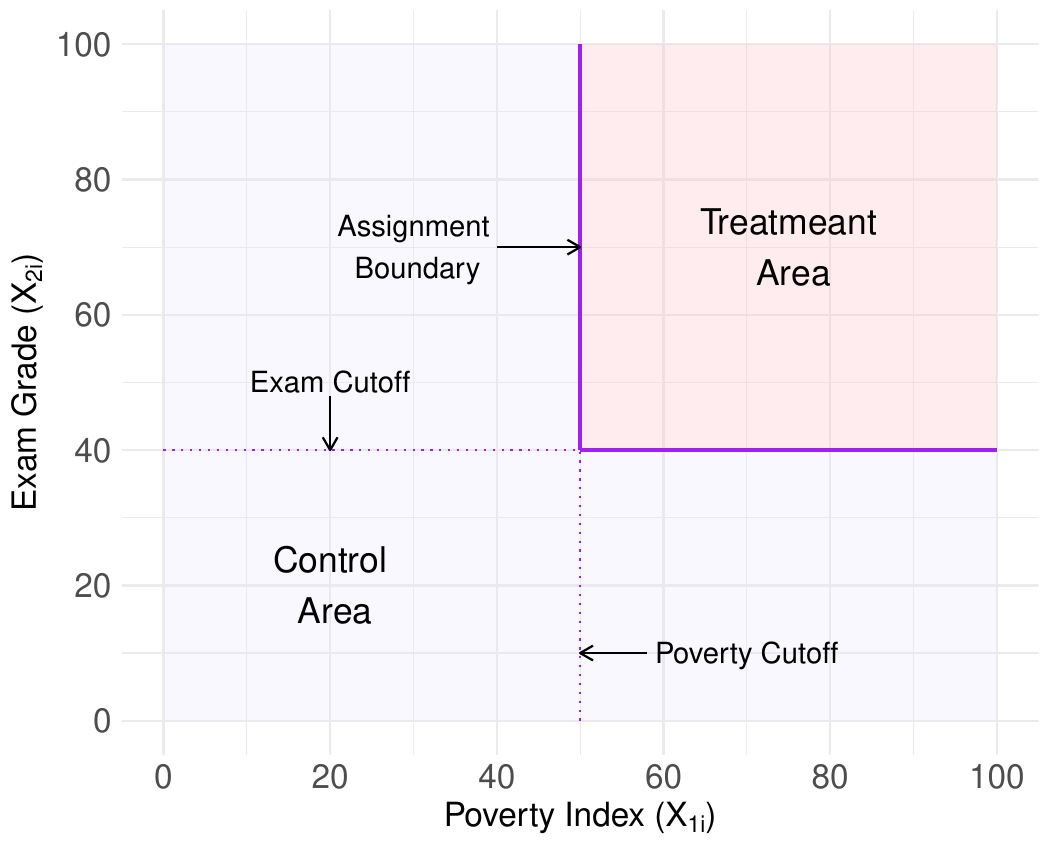}
        \caption{Two-Score RD Design.}
        \label{fig:fig1a}
    \end{subfigure}
    \quad
    \begin{subfigure}[b]{0.45\textwidth}
        \centering
        \includegraphics[width=\linewidth]{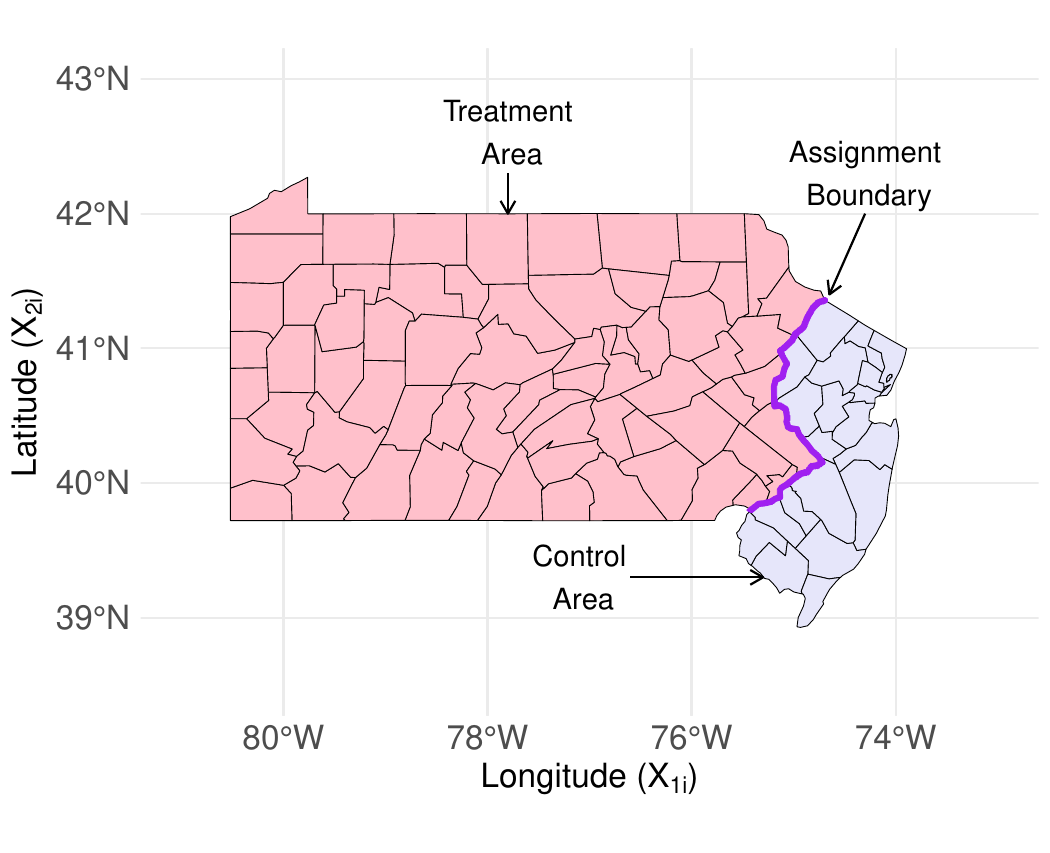}
        \caption{Geographic RD Design.}
        \label{fig:fig1b}
    \end{subfigure}

    \caption{Two Examples of Boundary Discontinuity Designs}
    \label{fig:fig1}
\end{figure}

Boundary discontinuity (BD) designs generalize the canonical RD design by allowing treatment assignment to be determined by a multidimensional score. The most common case is a bivariate score $\bX_i=(X_{1i},X_{2i})^\top$ and a one-dimensional assignment boundary $\B$ that separates treated and control regions. These designs are also called multi-score RD designs \citep{Papay-Willett-Murnane_2011_JoE,Reardon-Robinson_2012_JREE,Wong-Steiner-Cook_2013_JEBS}, and include geographic RD designs as a prominent special case \citep{Keele-Titiunik_2015_PA,Keele-Titiunik-Zubizarreta_2015_JRSSA,Keele-Titiunik_2016_PSRM,Keele-etal_2017_AIE,Galiani-McEwan-Quistorff_2017_AIE,Rischard-Branson-Miratrix-Bornn_2021_JASA,Diaz-Zubizarreta_2023_AOAS}. Figure~\ref{fig:fig1a} illustrates a two-score eligibility rule with a kinked boundary, while Figure~\ref{fig:fig1b} illustrates a generic geographic boundary. Empirical examples include education, taxation, political communication, labor markets, and place-based policy applications; see, among others, \citet{LondonoVelezRodriguezSanchez_2020_AEJ} and \citet{jardim2024local}. See \citet{Cattaneo-Idrobo-Titiunik_2020_CUP,Cattaneo-Idrobo-Titiunik_2024_CUP} for a two-part practical introductory monograph.

While classical RD designs based on a scalar score are well understood, BD designs have only recently received systematic methodological treatment. Empirical work has used different reductions of the boundary problem, often without formal guidance about which causal parameter is being estimated, how treatment-effect heterogeneity along the boundary should be summarized, or when inference is valid uniformly over many boundary locations. \citet{Cattaneo-Titiunik-Yu_2026_JASA} and \citet{Cattaneo-Titiunik-Yu_2026_JOE} address this gap by studying two leading approaches:
\begin{itemize}
    \item the \textit{location-based approach}, which uses the bivariate score directly and fits local polynomial regressions on the two sides of $\B$; and
    \item the \textit{distance-based approach}, which replaces the bivariate score with a scalar signed distance to each point on $\B$ and then applies univariate local polynomial methods.
\end{itemize}

\citet{Cattaneo-Titiunik-Yu_2026_JASA} develops location-based identification, estimation, bandwidth selection, and inference results, both pointwise for each point on $\B$ and uniformly over $\B$. \citet{Cattaneo-Titiunik-Yu_2026_JOE} develops the corresponding distance-based theory and clarifies when distance-to-boundary analyses recover the same causal targets. A central lesson is that distance-based methods can be useful and interpretable, but their validity depends more directly on the geometry of the boundary: kinks and other irregularities can generate additional bias even when the underlying bivariate conditional expectations are smooth. The location-based approach therefore provides the main benchmark whenever the bivariate score is available, while distance-based methods are valuable as a simpler implementation, a diagnostic, or a necessary alternative when only signed distances can be used.

This article introduces \pkg{rd2d}, software for \proglang{R}, \proglang{Python}, and \proglang{Stata} that implements these recent methodological developments for applied work. The package estimates the \textit{Boundary Average Treatment Effect Curve} (BATEC), the \textit{Weighted Boundary Average Treatment Effect} (WBATE), and the \textit{Largest Boundary Average Treatment Effect} (LBATE), together with fuzzy BATEC, fuzzy WBATE, and fuzzy LBATE for settings with imperfect treatment take-up. It provides automatic bandwidth selection, robust bias-corrected pointwise confidence intervals, uniform confidence bands, heteroskedasticity-robust and cluster-robust variance estimators, joint and separate fitting conventions, covariate-adjusted efficiency improvements, mass-point checks, covariance regularization for uniform inference, and derivative estimation for location-based methods. The robust bias-correction and coverage-error ideas build on \citet{Calonico-Cattaneo-Titiunik_2014_ECMA}, \citet{Calonico-Cattaneo-Farrell_2018_JASA}, \citet{Calonico-Cattaneo-Farrell_2020_ECTJ}, and \citet{Calonico-Cattaneo-Farrell_2022_Bernoulli}, adapted here to boundary-indexed local polynomial estimators. The \proglang{R} implementation centers on \code{rd2d()} and \code{rdbw2d()} for location-based analysis, and \code{rd2d.distance()} and \code{rdbw2d.distance()} for distance-based analysis; the \proglang{Python} and \proglang{Stata} implementations follow the same conceptual interface with language-specific syntax.

Two options deserve emphasis. First, the default option \code{fitmethod = "joint"} uses the treatment-interacted local regression representation when forming degrees-of-freedom adjustments, cluster-robust covariance matrices, and bandwidth-selection variance constants. This convention matters when a cluster, such as a school district, municipality, county, or tract group, can contain observations on both sides of the boundary: the cluster score should aggregate all local influence contributions from that cluster rather than treating the control and treated sides as independent samples. The option \code{fitmethod = "separate"} is retained for analyses that require the earlier side-specific convention. Second, the option \code{covs.eff} allows users to supply predetermined covariates for efficiency adjustment. These covariates enter with common coefficients across treatment sides and are propagated through both fixed-bandwidth estimation and automatic bandwidth selection.

The article is organized around the empirical steps in a BD analysis. We first clarify the causal target: a curve of local effects along the boundary, together with weighted and extremal summaries of that curve. We then show how the same targets are estimated using location-based and distance-based methods, how fuzzy designs decompose into reduced-form, first-stage, and Wald-ratio objects, and how software choices such as bandwidth selection, fitting convention, covariate adjustment, covariance storage, and boundary-grid construction affect empirical reporting. Throughout, an Opportunity Zones application provides the notation, code snippets, figures, and numerical output.

\subsection{Running Example: Opportunity Zones}\label{sec: Running Example}

We use the Opportunity Zones program as a running empirical example. The Tax Cuts and Jobs Act of 2017 allowed governors to nominate eligible low-income census tracts for Qualified Opportunity Zone designation. Following \citet{Corinth-Feldman_2024_JEP}, we construct a bivariate boundary discontinuity design in which eligibility is determined by two tract-level scores: a poverty-rate threshold and a median-family-income threshold. Actual designation is not mechanically assigned to all eligible tracts, which makes this a fuzzy BD design.

The public analysis dataset contains one observation per census tract. In the notation used below, $Y_i$ is the 2017--2019 change in $\log(1+\text{workplace jobs})$, $X_{1i}$ is the poverty rate minus 20, $X_{2i}$ is 80 minus the tract's median-family-income percentage, $T_i$ is low-income-community eligibility, and $W_i$ is Qualified Opportunity Zone designation. The assignment boundary is an L-shaped frontier with a kink at $(0,0)$. We evaluate treatment-effect curves at 31 boundary points: the kink and 15 points on each arm of the boundary, avoiding the far extremes where local effective sample sizes are thin.

\begin{figure}[t]
    \centering
    \includegraphics[width=0.74\linewidth]{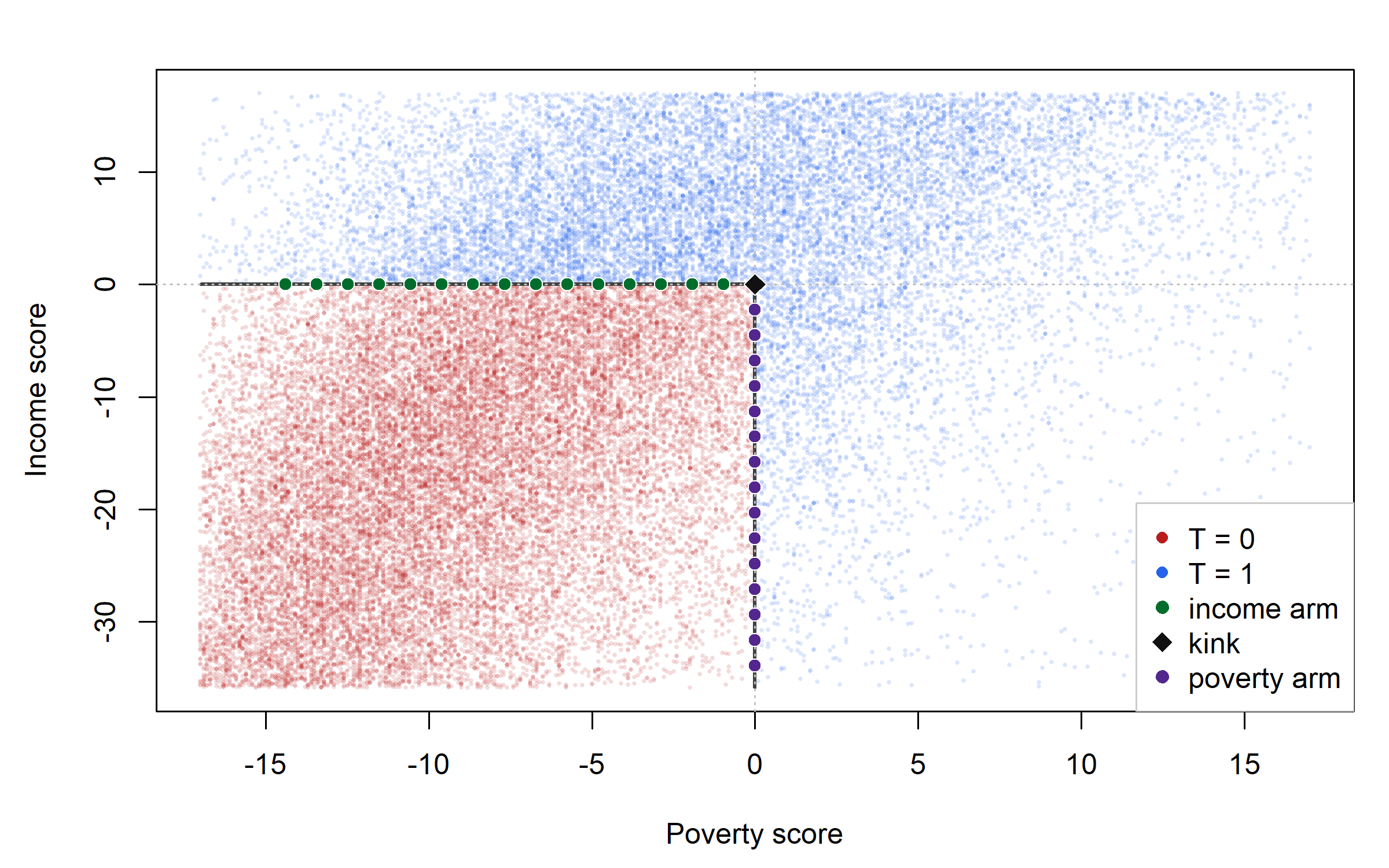}
    \caption{Opportunity Zones Running Example. The plot shows the final analysis sample, the low-income-community eligibility frontier, and the 31 evaluation points used in the empirical analysis.}
    \label{fig:empapp-scatter}
\end{figure}

\subsection{Organization and Background References}

The rest of the paper proceeds as follows. Section \ref{sec: Setup} introduces the BD setup, the main causal parameters, and their interpretation. Section \ref{sec: Location-Based Methods} discusses location-based estimation and inference, following the framework of \citet{Cattaneo-Titiunik-Yu_2026_JASA}. Section \ref{sec: Distance-Based Methods} discusses distance-based methods and the additional geometric considerations studied by \citet{Cattaneo-Titiunik-Yu_2026_JOE}. Section \ref{sec: Empirical Application} returns to the Opportunity Zones example and compares location-based and distance-based estimates. Section \ref{sec: Conclusion} concludes. Appendix \ref{sec: Multi Platform Syntax} summarizes the main \proglang{R}, \proglang{Python}, and \proglang{Stata} syntax. Replication codes, software documentation, and related information for \pkg{rd2d} are available at \url{https://rdpackages.github.io/}.

\section{Setup}\label{sec: Setup}

We employ standard potential outcomes notation \citep[and references therein]{Hernan-Robins_2020_Book}. Suppose that $(Y_i(0), Y_i(1), \bX_i^\top)^\top$, $i = 1,2,\dots, n$, is a random sample, where $Y_i(0)$ and $Y_i(1)$ denote the scalar potential outcomes for unit $i$ under control and treatment assignment, respectively. Units are assigned to control group or treatment group according to their bivariate location score $\bX_i= (X_{1i},X_{2i})^\top$ relative to a known one-dimensional boundary $\B$ splitting the support of $\bX_i$ in two disjoint regions: $\A_0$ denotes the control region, and $\A_1$ denotes the treatment region. Thus, $\B = \setbd(\A_0) \cap \setbd(\A_1)$, where $\setbd(\A_t)$ denotes the boundary of the set $\A_t$. We use the convention that boundary points belong to the treatment region, so $\B\subseteq\A_1$ and $\B\cap\A_0=\emptyset$. The observed response variable is $Y_i = \Indicator(\bX_i\in\A_0)Y_i(0) + \Indicator(\bX_i\in\A_1)Y_i(1)$, and $T_i = \Indicator(\bX_i \in \A_1)$ denotes treatment assignment. Figure \ref{fig:fig1} gives two graphical examples.

For implementation, the continuous assignment boundary $\B$ is first discretized into $J$ cutoff points $\bb=(\bb_1,\cdots,\bb_J)^\top$ with $\bb_j\in\B$ for all $j=1,\dots,J$. Then, the empirical analysis is conducted pointwise for each cutoff or uniformly over all cutoffs, employing either the bivariate location score $\bX_i$ directly, or an induced univariate distance to each cutoff point. See \citet[Section 5]{Cattaneo-Idrobo-Titiunik_2024_CUP} for an introductory discussion.

In the Opportunity Zones example, $\bX_i=(X_{1i},X_{2i})^\top$ contains the poverty and income scores shown in Figure~\ref{fig:empapp-scatter}; $\bb$ contains the 31 evaluation points on the L-shaped frontier; $T_i$ records eligibility; $W_i$ records actual designation; and $Y_i$ is early workplace-job growth. We use this notation in the code snippets below.

\subsection{Sharp BD Designs}

In a sharp BD design, assignment to the treatment region determines treatment exposure. The main causal parameter is the \textit{Boundary Average Treatment Effect Curve} (BATEC):
\begin{align*}
    \tau(\bx) = \E[Y_i(1) - Y_i(0) | \bX_i = \bx],
    \qquad \bx\in\B.
\end{align*}
The BATEC is a curve of local treatment effects indexed by points on the assignment boundary. It is useful when the policy rule creates many local comparisons rather than a single cutoff, because it allows researchers to inspect treatment-effect heterogeneity along the boundary before reporting lower-dimensional summaries.

In \pkg{rd2d}, the BATEC is obtained from the main output of a sharp BD design fit. The option \code{params.cov = "main"} stores the cross-boundary covariance matrix needed for uniform bands and aggregate summaries.
\begin{lstlisting}[style=Rstyle]
fit <- rd2d(Y = Y, X = X, assignment = T, b = b,
            params.cov = "main")
summary(fit, output = "main")
summary(fit, output = "main", cbands = "main")
\end{lstlisting}

The BATEC captures treatment effect heterogeneity along the assignment boundary, but researchers often want scalar summaries of that heterogeneity. For a weight function $w: \B \rightarrow [0,\infty)$, the \textit{Weighted Boundary Average Treatment Effect} (WBATE) is
\begin{align*}
    \tau_{\mathtt{WBATE}} = \frac{\int_\B \tau(\bx) w(\bx) \dif\bx}{\int_\B w(\bx) \dif\bx}.
\end{align*}
Different weights encode different policy or scientific targets. Equal weights over a pre-specified boundary grid produce a simple average of the BATEC values, while application-specific weights can emphasize boundary regions with more policy relevance or more population mass. A related scalar summary is the \textit{Largest Boundary Average Treatment Effect} (LBATE),
\begin{align*}
    \tau_{\mathtt{LBATE}} = \sup_{\bx\in\B}\tau(\bx),
\end{align*}
which summarizes the largest treatment effect along the boundary.

The WBATE aggregates the BATEC into a weighted average effect, while the LBATE reports the largest value of the curve. Thus, WBATE is an aggregation target and LBATE is an extremal heterogeneity target. In \pkg{rd2d}, WBATE estimation and inference are implemented via the optional argument \code{WBATE = w} in the \code{summary()} method for \code{rd2d} and \code{rd2d.distance} objects, while LBATE summaries are requested with \code{LBATE = TRUE}. Uniform bands, WBATE, and LBATE require the relevant cross-boundary covariance matrix to be stored.

\begin{lstlisting}[style=Rstyle]
w <- rep(1, nrow(b))
summary(fit, output = "main", cbands = "main",
        WBATE = w, LBATE = TRUE)
\end{lstlisting}

\subsection{Fuzzy BD Designs}

In a fuzzy BD design, the assignment rule changes the probability of treatment receipt but does not perfectly determine it. This is the structure of the Opportunity Zones example: eligibility $T_i$ changes the probability of designation $W_i$, but many eligible tracts are not designated and some policy choices are made after eligibility is determined. Let $T_i=\Indicator(\bX_i\in\A_1)$ denote assignment. Following \citet{Cattaneo-Titiunik-Yu_2026_JASA}, let $W_i(t)$ denote potential treatment status under assignment $t\in\{0,1\}$, and let $Y_i(t,w)$ denote the potential outcome under assignment $t$ and treatment status $w$. The observed treatment status and outcome are
\begin{align*}
    W_i
    &=
    \Indicator(\bX_i\in\A_0)W_i(0)
    +
    \Indicator(\bX_i\in\A_1)W_i(1),\\
    Y_i
    &=
    \Indicator(\bX_i\in\A_0)Y_i(0,W_i(0))
    +
    \Indicator(\bX_i\in\A_1)Y_i(1,W_i(1)).
\end{align*}
The reduced-form or intention-to-treat BATEC, abbreviated ITT, and the first-stage BATEC, abbreviated FS, are
\begin{align*}
    \tau_Y(\bx)
    &=
    \E[Y_i(1,W_i(1))-Y_i(0,W_i(0))\mid\bX_i=\bx],\\
    \tau_W(\bx)
    &=
    \E[W_i(1)-W_i(0)\mid\bX_i=\bx],
    \qquad \bx\in\B.
\end{align*}
Under the standard continuity conditions for fuzzy BD designs, these curves are identified by the observed boundary discontinuities
\begin{align*}
    \tau_A(\bx) = \lim_{\bu\to\bx,\bu\in\A_1}\E[A_i\mid\bX_i=\bu]
        - \lim_{\bu\to\bx,\bu\in\A_0}\E[A_i\mid\bX_i=\bu],
        \qquad A\in\{Y,W\}.
\end{align*}
When the first stage is nonzero and the standard fuzzy BD design assumptions hold locally, the fuzzy BATEC is the Wald ratio
\begin{align*}
    \zeta(\bx) = \frac{\tau_Y(\bx)}{\tau_W(\bx)},
    \qquad \bx\in\B.
\end{align*}
We refer to $\tau_Y(\bx)$ as the ITT BATEC, $\tau_W(\bx)$ as the FS BATEC, and $\zeta(\bx)$ as the fuzzy BATEC. The ITT curve captures heterogeneity in the effect of assignment on the outcome, the FS curve captures heterogeneity in compliance or take-up induced by assignment, and the fuzzy BATEC rescales the outcome discontinuity by the local first stage. The Wald ratio is a fuzzy BD estimand; interpreting it as a complier-average treatment effect additionally requires local exclusion, monotonicity, and a nonzero first stage near the boundary point.

Fuzzy analogues of WBATE and LBATE replace $\tau(\bx)$ with $\zeta(\bx)$. For a weight function $w: \B \rightarrow [0,\infty)$, the fuzzy WBATE is
\begin{align*}
    \zeta_{\mathtt{WBATE}} = \frac{\int_\B \zeta(\bx) w(\bx) \dif\bx}{\int_\B w(\bx) \dif\bx},
\end{align*}
and the fuzzy LBATE is
\begin{align*}
    \zeta_{\mathtt{LBATE}} = \sup_{\bx\in\B}\zeta(\bx).
\end{align*}
These summaries play the same role as in the sharp BD design: fuzzy WBATE aggregates the fuzzy BATEC over the boundary, while fuzzy LBATE reports the largest value of the fuzzy treatment-effect curve.

The package implements fuzzy BD designs by adding the treatment-receipt variable through \code{fuzzy}. The \code{output = "itt"} table reports the ITT BATEC, \code{output = "fs"} reports the FS BATEC, and \code{output = "main"} reports the fuzzy BATEC. The option \code{bwparam = "main"} targets automatic bandwidth selection to the fuzzy BATEC, while \code{bwparam = "itt"} targets the ITT BATEC.

\begin{lstlisting}[style=Rstyle]
fit_fuzzy <- rd2d(Y = Y, X = X, assignment = T, b = b,
                  fuzzy = W,
                  params.cov = c("main", "itt", "fs"))
summary(fit_fuzzy, output = "itt")
summary(fit_fuzzy, output = "fs")
summary(fit_fuzzy, output = "main")
\end{lstlisting}

Uniform bands and aggregate summaries for fuzzy BD designs are requested output by output. For example, the following calls report the ITT BATEC, FS BATEC, and fuzzy BATEC with uniform bands; the final call also reports fuzzy WBATE and fuzzy LBATE.
\begin{lstlisting}[style=Rstyle]
w <- rep(1, nrow(b))
summary(fit_fuzzy, output = "itt", cbands = "itt")
summary(fit_fuzzy, output = "fs", cbands = "fs")
summary(fit_fuzzy, output = "main", cbands = "main",
        WBATE = w, LBATE = TRUE)
\end{lstlisting}

\section{Location-Based Methods}\label{sec: Location-Based Methods}

Location-based methods use the bivariate score $\bX_i$ directly and estimate side-specific local polynomial approximations at the boundary. In implementation these side-specific fits are represented through a joint treatment-interacted regression, which is equivalent for point estimation and is used for covariance calculations. Under the usual continuity assumptions invoked in canonical RD designs \citep{Hahn-Todd-vanderKlaauw_2001_ECMA}, the BATEC in a sharp BD design is identified by
\begin{align*}
    \tau(\bx) = \mu_1(\bx) - \mu_0(\bx),
    \qquad \bx\in\B,
\end{align*}
where $\mu_t(\bx)=\E[Y_i(t)|\bX_i=\bx]=\E[Y_i|\bX_i=\bx,T_i=t]$, $t\in\{0,1\}$, are smooth functions on their corresponding sides of the boundary. This is the standard continuity argument for RD designs, adapted to a bivariate score and a one-dimensional assignment boundary \citep{Papay-Willett-Murnane_2011_JoE,Reardon-Robinson_2012_JREE,Wong-Steiner-Cook_2013_JEBS,Keele-Titiunik_2015_PA}. Identification of WBATE and LBATE follows from identification of $\tau(\bx)$ over $\B$. In fuzzy BD designs, the same continuity argument identifies the reduced-form BATEC $\tau_Y(\bx)$ and the first-stage BATEC $\tau_W(\bx)$; when $\tau_W(\bx)$ is bounded away from zero on the target boundary, the fuzzy BATEC $\zeta(\bx)=\tau_Y(\bx)/\tau_W(\bx)$ and its WBATE and LBATE analogues are identified, with the complier-average interpretation requiring the additional conditions discussed in Section~\ref{sec: Setup}. Following \citet{Cattaneo-Titiunik-Yu_2026_JASA}, this section focuses on estimation, bandwidth selection, and uncertainty quantification for location-based methods.

\subsection{Treatment Effect Estimation: Methods}\label{sec: Location Estimation Methods}

For each $\bx\in\B$, the location-based estimator can be written as a single local polynomial regression with treatment-side interactions. Let $\br_p(\bu)$ collect all monomials $u_1^{a_1}u_2^{a_2}$ with $a_1+a_2\leq p$, starting with the constant term, and let $\mathfrak{p}_p=(2+p)(1+p)/2$. With $T_i=\Indicator(\bX_i\in\A_1)$, define the augmented basis
\[
    \br_{p,h}(\bu;T)
    =
    \begin{pmatrix}
        \br_p(\bu/h) &
        T\br_p(\bu/h)
    \end{pmatrix}^\top.
\]
For any multi-index $\bnu$ with $|\bnu|\le p$, let $\be_{\bnu}$ denote the length-$2\mathfrak{p}_p$ vector that selects the coefficient on the treatment-interaction monomial indexed by $\bnu$ in $\br_{p,h}(\bu;T)$; in particular, $\be_{\mathbf{0}}$ selects the treatment-interaction intercept. The local polynomial estimator is
\begin{align*}
    \widehat{\bdelta}(\bx) =
    \argmin_{\bdelta\in\mathbb{R}^{2\mathfrak{p}_p}}
    \frac{1}{n}\sum_{i=1}^n
        \big(Y_i-\br_{p,h}(\bX_i-\bx; T_i)^\top \bdelta\big)^2
        K_h(\bX_i-\bx),
\end{align*}
where $K_h(\bu)=K(u_1/h,u_2/h)/h^2$. The common scalar bandwidth $h$ is used for exposition; in implementation the score coordinates can be standardized, and separate bandwidths can be used by coordinate and by treatment side.
For interpretation, decompose the coefficient vector as $\widehat{\bdelta}(\bx)=(\widehat{\bdelta}_0(\bx)^\top,\widehat{\bdelta}_1(\bx)^\top)^\top$ with $\widehat{\bdelta}_0(\bx),\widehat{\bdelta}_1(\bx)\in\mathbb{R}^{\mathfrak{p}_p}$. Then, $\widehat{\bdelta}_0(\bx)$ describes the control-side local polynomial, while $\widehat{\bdelta}_1(\bx)$ describes the effects due to treatment assignment. Thus the BATEC estimator is
\begin{align*}
    \widehat{\tau}(\bx)
    = \be_{\mathbf{0}}^{\top}\widehat{\bdelta}(\bx),
    \qquad \bx\in\B.
\end{align*}
This interaction representation is useful for software implementation because additional terms, such as predetermined covariates, can be added to the same weighted regression while leaving the treatment-side intercept interaction as the treatment-effect parameter.

The leading MSE terms depend on the local design, conditional variance, and curvature of the regression functions on each side of the boundary. Let $\sigma_t^2(\bx)=\Var[Y_i(t)|\bX_i=\bx]$. Using the augmented basis, define
\begin{align*}
    \bGamma_{\bx}
    &=
    \E\!\left[
    \br_{p,h}(\bX_i-\bx;T_i)\br_{p,h}(\bX_i-\bx;T_i)^\top
    K_h(\bX_i-\bx)
    \right],\\
    \bSigma_{\bx_1,\bx_2}
    &=
    h^2\E\!\left[\br_{p,h}(\bX_i-\bx_1;T_i)\br_{p,h}(\bX_i-\bx_2;T_i)^\top K_h(\bX_i-\bx_1)K_h(\bX_i-\bx_2)\sigma_{T_i}^2(\bX_i)\right].
\end{align*}
Then the variance constant for the BATEC is
\[
    V_{\bx}
    =
    \be_{\mathbf{0}}^{\top}\bGamma_{\bx}^{-1}
    \bSigma_{\bx,\bx}
    \bGamma_{\bx}^{-1}\be_{\mathbf{0}}.
\]
Using multi-index notation, the leading bias is
\begin{align*}
    B_{\bx}
    =
    \be_{\mathbf{0}}^{\top}\bGamma_{\bx}^{-1}
    \sum_{t=0}^{1}\sum_{|\bk|=p+1}
    \frac{\mu_t^{(\bk)}(\bx)}{\bk!}
    \E\!\left[\br_{p,h}(\bX_i-\bx;T_i)\left((\bX_i-\bx)/h\right)^{\bk}K_h(\bX_i-\bx)\Indicator(\bX_i\in\A_t)\right].
\end{align*}

Under the regularity conditions in \citet{Cattaneo-Titiunik-Yu_2026_JASA}, $\widehat{\tau}(\bx)$ is pointwise and uniformly consistent for $\tau(\bx)$ as $h\to0$ and $nh^2$ grows. The pointwise conditional MSE expansion has the usual bias-variance form,
\begin{align*}
    \E[(\widehat{\tau}(\bx)-\tau(\bx))^2|\bX]
    \approx_\P
    h^{2p+2}B_{\bx}^2+\frac{1}{nh^2}V_{\bx},
    \qquad \bx\in\B,
\end{align*}
with the leading bias and variance constants defined above. Integrating the same expansion over $\B$ gives
\begin{align*}
    \int_{\B}\E[(\widehat{\tau}(\bx)-\tau(\bx))^2|\bX]w(\bx)\dif\bx
    \approx_\P
    h^{2p+2}\int_{\B}B_{\bx}^2w(\bx)\dif\bx
    +\frac{1}{nh^2}\int_{\B}V_{\bx}w(\bx)\dif\bx.
\end{align*}
Ignoring higher-order terms gives the infeasible pointwise MSE and integrated MSE bandwidths
\begin{align*}
    h_{\mathtt{MSE},\bx}
    = \left(\frac{2V_{\bx}}{(2p+2)B_{\bx}^2}\frac{1}{n}\right)^{1/(2p+4)},
    \qquad
    h_{\mathtt{IMSE}}
    = \left(\frac{2\int_{\B}V_{\bx}w(\bx)\dif\bx}
    {(2p+2)\int_{\B}B_{\bx}^2w(\bx)\dif\bx}\frac{1}{n}\right)^{1/(2p+4)}.
\end{align*}
The pointwise selector targets a particular boundary location, while the integrated selector targets the full boundary curve or a chosen subset of it.

\subsection{Treatment Effect Estimation: Implementation}\label{sec: Location Estimation Implementation}

Location-based estimation is implemented by \code{rd2d()}, and bandwidth selection can be called separately through \code{rdbw2d()}. In the running example, \code{X} contains the poverty and income scores and \code{b} contains the boundary grid. A basic sharp BD design fit is
\begin{lstlisting}[style=Rstyle]
fit <- rd2d(Y = Y, X = X, assignment = T, b = b,
            params.cov = "main")
summary(fit, output = "main")
\end{lstlisting}
The resulting \code{summary()} table has one row per boundary evaluation point. It reports the order-$p$ point estimate and standard error, the order-$q$ bias-corrected estimate and standard error used for RBC inference, the associated test statistic, $p$-value, pointwise confidence interval, selected bandwidths, and side-specific effective sample sizes. When \code{cbands}, \code{WBATE}, or \code{LBATE} are requested, the same summary call appends uniform confidence-band columns or aggregate rows. The option \code{params.cov = "main"} stores the covariance matrix for the main sharp BD design estimand, which is needed for uniform bands and aggregate boundary summaries.

The implementation can report up to four bandwidths: $h_{01}$ and $h_{02}$ for the first and second score coordinates on the control side, and $h_{11}$ and $h_{12}$ for the corresponding treatment-side coordinates. The option \code{stdvars=TRUE}, the default, computes bandwidths after standardizing each score coordinate and then rescales them to the original units. The main \code{bwselect} choices are:
\begin{itemize}
    \item \code{mserd} and \code{cerrd}: pointwise MSE and CER selectors with common bandwidths across treatment sides.
    \item \code{imserd} and \code{icerrd}: integrated MSE and CER selectors with common bandwidths across treatment sides.
    \item \code{msetwo} and \code{certwo}: pointwise MSE and CER selectors with separate control and treatment bandwidths.
    \item \code{imsetwo} and \code{icertwo}: integrated MSE and CER selectors with separate control and treatment bandwidths.
\end{itemize}
Unless overridden by the user, \code{rd2d()} uses a triangular product kernel, $p=1$ for point estimation, $q=p+1$ for RBC inference, \code{vce = "hc1"}, \code{fitmethod = "joint"}, \code{masspoints = "check"}, \code{level = 95}, and automatic bandwidth selection with \code{bwselect = "mserd"} and \code{method = "dpi"}. The option \code{method = "rot"} requests a rule-of-thumb implementation. If the user supplies \code{h}, the reported selector is \code{"user provided"}.

The bandwidth selector can also be inspected before estimation:
\begin{lstlisting}[style=Rstyle]
bw <- rdbw2d(Y = Y, X = X, assignment = T, b = b,
             bwselect = "imserd", stdvars = TRUE)
summary(bw)

fit_imse <- rd2d(Y = Y, X = X, assignment = T, b = b,
                 bwselect = "imserd", params.cov = "main")
\end{lstlisting}

The variance estimator is controlled by \code{vce} and \code{cluster}. When \code{cluster = NULL}, the package reports heteroskedasticity-consistent standard errors; when \code{cluster} is supplied, it reports cluster-robust standard errors. The \code{vce} choices are:
\begin{itemize}
    \item \code{hc0}: plug-in residual variance with no degrees-of-freedom or leverage adjustment.
    \item \code{hc1}: degrees-of-freedom adjustment; this is the default.
    \item \code{hc2}: leverage-adjusted residual weighting.
    \item \code{hc3}: squared leverage-adjusted residual weighting.
\end{itemize}
The fitting convention used for variance estimation and bandwidth selection is controlled by \code{fitmethod}. The default, \code{fitmethod = "joint"}, implements the treatment-interacted regression described above. In designs without covariates, the joint and separate conventions give the same local polynomial point estimates, but they can give different standard errors and automatic bandwidths because the finite-sample degrees-of-freedom factors and cluster scores are formed differently. The joint convention is particularly relevant for clustered data because a cluster can contribute observations to both sides of the boundary and to several nearby boundary points. In that case, cluster-robust inference should add the control-side and treatment-side influence contributions before forming the cluster covariance. The option \code{fitmethod = "separate"} uses the earlier side-specific convention and is useful for replication or sensitivity checks.

\begin{lstlisting}[style=Rstyle]
# cluster_id can be, for example, a school, county, or tract group.
fit_cl <- rd2d(Y = Y, X = X, assignment = T, b = b,
               cluster = cluster_id,
               fitmethod = "joint",
               params.cov = "main")

fit_cl_separate <- rd2d(Y = Y, X = X, assignment = T, b = b,
                        cluster = cluster_id,
                        fitmethod = "separate")
\end{lstlisting}
The plug-in bandwidth rules also estimate score variances for standardization, preliminary residuals for variance constants, and higher-order derivatives of the conditional mean for bias constants. The preliminary bandwidths used for these quantities follow the direct plug-in procedure described in \citet{Cattaneo-Titiunik-Yu_2026_JASA} and its supplemental appendix.

\subsection{Statistical Inference: Methods}\label{sec: Location Inference Methods}

For a chosen bandwidth and polynomial order, define the feasible covariance kernel $\widehat{\Omega}_{\bx_1,\bx_2}$. For example, in the sharp BD design,
\begin{align*}
    \widehat{\Omega}_{\bx_1,\bx_2}
    =
    \frac{1}{nh^2}\be_{\mathbf{0}}^{\top}
    \widehat{\bGamma}_{\bx_1}^{-1}
    \widehat{\bSigma}_{\bx_1,\bx_2}
    \widehat{\bGamma}_{\bx_2}^{-1}
    \be_{\mathbf{0}}.
\end{align*}
Here
\[
    \widehat{\bGamma}_{\bx}
    =
    \frac{1}{n}\sum_{i=1}^n
    \br_{p,h}(\bX_i-\bx;T_i)\br_{p,h}(\bX_i-\bx;T_i)^\top
    K_h(\bX_i-\bx),
\]
and
\[
    \widehat{\bSigma}_{\bx_1,\bx_2}=\frac{h^2}{n}\sum_{i=1}^{n}\big[\br_{p,h}(\bX_i-\bx_1;T_i)\br_{p,h}(\bX_i-\bx_2;T_i)^\top K_h(\bX_i-\bx_1)K_h(\bX_i-\bx_2)\widehat{\varepsilon}_i(\bx_1)\widehat{\varepsilon}_i(\bx_2)\big],
\]
where $\widehat{\varepsilon}_i(\bx)$ is the residual from the augmented local polynomial fit at $\bx$. These objects are also what make uniform bands and aggregate standard errors depend on covariance across boundary points, not only on pointwise standard errors.
The pointwise Wald interval is
\begin{align*}
    \widehat{\CI}_{\alpha}(\bx)
    =
    \left[
        \widehat{\tau}(\bx)-q_{\alpha}\widehat{\Omega}_{\bx,\bx}^{1/2},
        \widehat{\tau}(\bx)+q_{\alpha}\widehat{\Omega}_{\bx,\bx}^{1/2}
    \right],
    \qquad \bx\in\B.
\end{align*}
For pointwise inference, $q_{\alpha}=\Phi^{-1}(1-\alpha/2)$. For uniform inference over $\B$, the critical value is chosen to approximate the distribution of $\sup_{\bx\in\B}|\widehat{Z}(\bx)|$, where $\widehat{Z}(\bx)$ is a conditionally mean-zero Gaussian process with covariance induced by $\widehat{\Omega}_{\bx_1,\bx_2}$. In practice, this process is simulated on the boundary grid $\bb_1,\ldots,\bb_J$.

Because MSE-optimal bandwidths leave non-negligible smoothing bias, \pkg{rd2d} implements robust bias-corrected (RBC) inference, following the local-polynomial RD literature \citep{Calonico-Cattaneo-Titiunik_2014_ECMA,Calonico-Cattaneo-Farrell_2018_JASA,Calonico-Cattaneo-Farrell_2020_ECTJ,Calonico-Cattaneo-Farrell_2022_Bernoulli}. With a $p$th order polynomial used for point estimation and a $q$th order polynomial, typically $q=p+1$, used for inference, let $\widehat{\tau}_{\mathtt{bc}}(\bx)$ denote the bias-corrected estimator and let $\widehat{\Omega}_{\mathtt{bc},\bx_1,\bx_2}$ denote the corresponding covariance kernel, including the extra variability from estimating the leading bias. The RBC interval is
\begin{align*}
    \widehat{\CI}_{\mathtt{bc},\alpha}(\bx)
    =
    \left[
        \widehat{\tau}_{\mathtt{bc}}(\bx)-q_{\alpha}\widehat{\Omega}_{\mathtt{bc},\bx,\bx}^{1/2},
        \widehat{\tau}_{\mathtt{bc}}(\bx)+q_{\alpha}\widehat{\Omega}_{\mathtt{bc},\bx,\bx}^{1/2}
    \right].
\end{align*}
The resulting intervals and bands are generally not centered at the original BATEC point estimator $\widehat{\tau}(\bx)$.

\subsection{Statistical Inference: Implementation}\label{sec: Location Inference Implementation}

Pointwise intervals are reported by default through the \code{summary()} method. Uniform bands are requested with \code{cbands}; the corresponding covariance matrix must be stored when the model is fit.
\begin{lstlisting}[style=Rstyle]
fit <- rd2d(Y = Y, X = X, assignment = T, b = b,
            params.cov = "main")
summary(fit, output = "main")
summary(fit, output = "main", cbands = "main")
\end{lstlisting}
The default polynomial orders are $p=1$ for point estimation and $q=2$ for RBC inference. The option \code{level} controls the nominal coverage level, and \code{repp} controls the number of Gaussian simulations used for uniform critical values. Finite-sample covariance regularization may be used when the simulated Gaussian covariance matrix is nearly singular, a situation that can arise with closely spaced boundary grid points or weak local support.

\subsection{WBATE and LBATE: Estimation and Inference}\label{sec: Location Aggregate Inference}

The WBATE is fit by plugging the estimated BATEC into the boundary integral. With normalized weights $\int_{\B}w(\bx)\dif\bx=1$,
\begin{align*}
    \widehat{\tau}_{\mathtt{WBATE}}
    =
    \int_{\B}\widehat{\tau}(\bx)w(\bx)\dif\bx.
\end{align*}
On a grid, this is approximated by $\sum_{j=1}^J\omega_j\widehat{\tau}(\bb_j)w(\bb_j)$, with quadrature weights $\omega_j$ normalized so the weighted sum integrates to one. The WBATE covariance is the corresponding double integral of the BATEC covariance kernel,
\begin{align*}
    \widehat{\Omega}_{\mathtt{WBATE}}
    =
    \int_{\B}\int_{\B}
    \widehat{\Omega}_{\bx_1,\bx_2}
    w(\bx_1)w(\bx_2)\dif\bx_1\dif\bx_2,
\end{align*}
or its grid analogue. Aggregation changes the effective variance order: the pointwise BATEC estimator has variance of order $(nh^2)^{-1}$, while the WBATE has variance of order $(nh)^{-1}$ because only boundary locations within an $O(h)$ neighborhood have non-negligible covariance. Thus, the WBATE behaves like a one-dimensional nonparametric functional even though each local fit uses a two-dimensional score. Inference uses a normal critical value, with RBC implemented by applying the same plug-in formula to the bias-corrected BATEC and its covariance.

The LBATE is the supremum of the BATEC, with plug-in estimator
\begin{align*}
    \widehat{\tau}_{\mathtt{LBATE}}
    =
    \sup_{\bx\in\B}\widehat{\tau}(\bx)
    \approx
    \max_{1\leq j\leq J}\widehat{\tau}(\bb_j).
\end{align*}
Because LBATE is an extreme functional, \citet{Cattaneo-Titiunik-Yu_2026_JASA} recommends reporting it together with the full BATEC and WBATE. Its confidence interval is obtained by projecting the uniform confidence band for $\tau(\bx)$:
\begin{align*}
    \widehat{\CI}_{\alpha,\mathtt{LBATE}}
    =
    \left[
    \sup_{\bx\in\B}\left(\widehat{\tau}(\bx)-q_{\alpha}\widehat{\Omega}_{\bx,\bx}^{1/2}\right),
    \sup_{\bx\in\B}\left(\widehat{\tau}(\bx)+q_{\alpha}\widehat{\Omega}_{\bx,\bx}^{1/2}\right)
    \right],
\end{align*}
using the uniform critical value $q_{\alpha}$. The interval is generally conservative because it inherits simultaneous coverage for the full BATEC.

In the software, both summaries are produced from the stored covariance matrix:
\begin{lstlisting}[style=Rstyle]
w <- rep(1, nrow(b))
summary(fit, output = "main", cbands = "main",
        WBATE = w, LBATE = TRUE)
\end{lstlisting}

\subsection{Imperfect Compliance}\label{sec: Location Fuzzy Methods}

For fuzzy BD designs, the same local polynomial machinery is applied to the outcome $Y_i$ and the treatment status $W_i$. Let $\widehat{\tau}_{Y}(\bx)$ and $\widehat{\tau}_{W}(\bx)$ denote the reduced-form and first-stage BATEC estimators. The fuzzy BATEC estimator is
\begin{align*}
    \widehat{\zeta}(\bx)
    =
    \frac{\widehat{\tau}_{Y}(\bx)}{\widehat{\tau}_{W}(\bx)}.
\end{align*}
Here $\tau_Y(\bx)$ is the ITT BATEC and $\tau_W(\bx)$ is the FS BATEC. The ratio $\zeta(\bx)$ is the fuzzy BATEC parameter; it has a local complier-average causal interpretation only under additional fuzzy BD design assumptions such as a local exclusion restriction, monotonicity, and a nonzero first stage. For related discussion of causal identification in fuzzy multidimensional RD designs, see \citet{choi2023complier}, \citet{schwarz2025effect}, \citet{jiang2026extrapolating}, and \citet{Cattaneo-Keele-Titiunik-VazquezBare_2016_JOP,Cattaneo-Keele-Titiunik-VazquezBare_2021_JASA}; for univariate fuzzy RD designs, see \citet{Arai-etal_2022_QE} and references therein.

\citet{Cattaneo-Titiunik-Yu_2026_JASA} uses the linearization
\begin{align*}
    \widehat{\zeta}(\bx)-\zeta(\bx)
    =
    \mathfrak{w}(\bx)^\top\mathfrak{T}(\bx)
    +\mathfrak{R}_{\mathtt{F}}(\bx),
    \qquad
    \mathfrak{w}(\bx)
    =
    \left[
    \frac{1}{\tau_W(\bx)},
    -\frac{\tau_Y(\bx)}{\tau_W(\bx)^2}
    \right]^\top,
\end{align*}
where $\mathfrak{T}(\bx)=(\widehat{\tau}_Y(\bx)-\tau_Y(\bx),\widehat{\tau}_W(\bx)-\tau_W(\bx))^\top$. The remainder $\mathfrak{R}_{\mathtt{F}}(\bx)$ is negligible when the first stage is uniformly bounded away from zero and the reduced-form and first-stage estimators are uniformly consistent. Thus, estimation and inference for fuzzy BATEC, fuzzy WBATE, and fuzzy LBATE follow by applying the sharp BD design results to the linearized combination $\mathfrak{w}(\bx)^\top\mathfrak{T}(\bx)$.

The fuzzy covariance kernel is
\begin{align*}
    \widehat{\Omega}_{\mathtt{F},\bx_1,\bx_2}
    =
    \widehat{\mathfrak{w}}(\bx_1)^\top
    \begin{pmatrix}
        \widehat{\Omega}_{Y,Y,\bx_1,\bx_2} & \widehat{\Omega}_{Y,W,\bx_1,\bx_2}\\
        \widehat{\Omega}_{W,Y,\bx_1,\bx_2} & \widehat{\Omega}_{W,W,\bx_1,\bx_2}
    \end{pmatrix}
    \widehat{\mathfrak{w}}(\bx_2),
\end{align*}
where $\widehat{\mathfrak{w}}(\bx)$ replaces $\tau_Y(\bx)$ and $\tau_W(\bx)$ by their local polynomial estimates. Fuzzy BATEC intervals and bands use $\widehat{\Omega}_{\mathtt{F},\bx_1,\bx_2}$ in place of the sharp BD design covariance kernel. Fuzzy WBATE uses
\[
    \widehat{\zeta}_{\mathtt{WBATE}}
    =
    \int_{\B}\widehat{\zeta}(\bx)w(\bx)\dif\bx
    \quad\text{and}\quad
    \widehat{\Omega}_{\mathtt{F,WBATE}}
    =
    \int_{\B}\int_{\B}
    \widehat{\Omega}_{\mathtt{F},\bx_1,\bx_2}
    w(\bx_1)w(\bx_2)\dif\bx_1\dif\bx_2,
\]
with a normal critical value. Fuzzy LBATE uses $\widehat{\zeta}_{\mathtt{LBATE}}=\sup_{\bx\in\B}\widehat{\zeta}(\bx)$ and projects the fuzzy BATEC uniform band in the same way as for LBATE in a sharp BD design.

In \pkg{rd2d}, fuzzy BD designs are fit by supplying \code{fuzzy = W}. For Opportunity Zones, this means that the first-stage output estimates the effect of eligibility on designation, while the ITT output estimates the effect of eligibility on early workplace-job growth. The output \code{"itt"} reports $\widehat{\tau}_Y(\bx)$, \code{"fs"} reports $\widehat{\tau}_W(\bx)$, and \code{"main"} reports $\widehat{\zeta}(\bx)$. The option \code{bwparam = "main"} targets the fuzzy BATEC, while \code{bwparam = "itt"} targets the reduced-form BATEC.
\begin{lstlisting}[style=Rstyle]
fit_fuzzy <- rd2d(Y = Y, X = X, assignment = T, b = b,
                  fuzzy = W,
                  params.cov = c("main", "itt", "fs"))
w <- rep(1, nrow(b))
summary(fit_fuzzy, output = "itt", cbands = "itt")
summary(fit_fuzzy, output = "fs", cbands = "fs")
summary(fit_fuzzy, output = "main", cbands = "main",
        WBATE = w, LBATE = TRUE)
\end{lstlisting}

\subsection{Regularization Strategies and Implementation Choices}\label{sec: Implementation and Regularization Details}

The package \pkg{rd2d} implements several numerical safeguards and reporting choices that are important in empirical applications. These choices do not change the target parameters; they stabilize the finite-sample implementation of BATEC, WBATE, LBATE, and their fuzzy analogues, and make the resulting diagnostics easier to interpret.

\begin{itemize}
    \item \textit{Boundary grid and target set}. Uniform bands, WBATE, and LBATE are computed on the user-supplied boundary grid \code{b}. For a piecewise linear boundary, a natural default is to place grid points approximately equally spaced in arc length, while forcing kinks, corners, or substantively important boundary points to be included. If inference is desired only on a trimmed subset $\B_0\subseteq\B$, for example to avoid endpoints or poorly supported corners, the grid should represent $\B_0$ and the reported estimands should be interpreted as applying to that trimmed target set.

    \item \textit{Grid sensitivity}. Since the theoretical uniform band is indexed by the continuum $\bx\in\B$ but the software simulates the process on a finite grid, the grid should be fine enough that the discretized supremum is stable. A useful diagnostic is to refine the grid and check whether the simulated critical value, the confidence band, and aggregate summaries are essentially unchanged. In applications with heterogeneous effects or irregular boundaries, this check is more informative than relying on a fixed number of grid points.

    \item \textit{Bandwidth choices along the boundary}. Pointwise selectors target a particular boundary location, whereas integrated selectors target the full boundary curve or a chosen subset. For uniform inference and graphical displays, a common boundary-wide bandwidth often gives smoother covariance estimates and simpler interpretation. Point-specific bandwidths can also be used, but the covariance matrix used for uniform bands and aggregate summaries must be computed using the same bandwidth choices. The default standardization option, \code{stdvars = TRUE}, helps make scalar bandwidths comparable when the two score coordinates have different units.

    \item \textit{Joint versus separate fitting}. The option \code{fitmethod = "joint"} is the default because it matches the treatment-interacted regression used to define the estimator and because it aggregates cluster scores across both sides of the boundary. This distinction matters when the same cluster can contain control and treated units near a boundary point. The option \code{fitmethod = "separate"} is available when users want side-specific finite-sample adjustments for comparison with earlier implementations.

    \item \textit{Covariate efficiency adjustment}. The option \code{covs.eff} supplies predetermined covariates used to improve precision without changing the treatment-effect estimand. The same adjustment is used in automatic bandwidth selection, pointwise inference, and covariance matrices for uniform bands and aggregate summaries. The safeguards \code{covs.drop} and \code{covs.tol} control how redundant covariate columns are handled after residualization.

    \item \textit{Small bias regularization}. Approximate MSE and IMSE bandwidth choices require nonzero leading bias terms. A very small estimated bias can therefore produce an excessively large bandwidth. To avoid this problem, the software adds a regularization term to the estimated bias denominator:
    \begin{align*}
        h_{\mathtt{MSE},\bx}
        =
        \left(
        \frac{2\widehat{V}_{\bx}}
        {(2p+2)(\widehat{B}_{\bx}^2+s\cdot\Var[\widehat{B}_{\bx}])}
        \frac{1}{n}
        \right)^{1/(2p+4)}.
    \end{align*}
    The analogous IMSE selector regularizes $\int_{\B}\widehat{B}_{\bx}^2w(\bx)\dif\bx$ by the integrated variance of the bias estimator. The factor $s$ is controlled by \code{scaleregul}; its default is $3$ when bandwidth selection is called through \code{rd2d()} and $1$ in standalone calls to \code{rdbw2d()}.

    \item \textit{Minimum sample size}. The option \code{bwcheck} enlarges the selected or user-provided bandwidth until a minimum number of observations is included in the estimation region on each side of the cutoff. For \code{rd2d()}, the default is $\text{\code{bwcheck}}=50+p+1$; standalone calls to \code{rdbw2d()} use $\text{\code{bwcheck}}=20$.

    \item \textit{Mass points in $\bX_i$}. The option \code{masspoints = "check"} reports the number of unique score values and warns when duplication is substantial. With \code{masspoints = "adjust"}, bandwidths are regularized so that kernels contain a minimum number of unique observations. With \code{masspoints = "off"}, the potential presence of mass points is ignored.

    \item \textit{Covariance regularization for uniform bands}. The estimated covariance matrix is converted to a correlation matrix before Gaussian simulation. In finite samples this matrix can be nearly singular or slightly fail to be positive semidefinite because of numerical precision, closely spaced grid points, or weak local support. The implementation symmetrizes the matrix, truncates eigenvalues below a small tolerance, rescales the diagonal entries to one, and warns when the adjustment is nonnegligible. This eigenvalue regularization follows the same logic as other covariance regularization methods for uniform inference \citep{Cattaneo-Feng-Underwood_2024_JASA}.

    \item \textit{Gaussian simulation error}. The option \code{repp} controls the number of Gaussian simulations used for uniform critical values and LBATE inference. The default is \code{repp = 1000} in the summary methods. When the critical value is central to the empirical conclusion, increasing \code{repp} and checking that the reported band is stable is a simple numerical diagnostic.

    \item \textit{Local support and design stability}. Boundary points with little one-sided support, very unbalanced kernel mass across the two sides, or poorly conditioned local design matrices can lead to unstable BATEC estimates and covariance estimates. The \code{bwcheck} and \code{masspoints} options address two common symptoms, but researchers should also inspect the score scatterplot, side-specific local sample sizes, and sensitivity to trimming weakly supported regions of the boundary. These checks matter most near kinks, corners, and endpoints.

    \item \textit{Reporting practice}. Empirical analyses should report the target boundary set, the evaluation grid, the bandwidth selector, the fitting convention, any clustering or covariate adjustment, whether covariance regularization produced a warning, and the simulation count used for uniform bands. Basic sensitivity checks with alternative grid sizes, bandwidth choices, and, when relevant, boundary trims help separate genuine treatment-effect heterogeneity from discretization, weak support, or numerical instability.
\end{itemize}

The empirical implementation can be organized as follows. First, construct and plot the boundary grid, making sure that kinks, endpoints, and substantively important locations are represented. Second, estimate the location-based BATEC with covariance storage for the outputs that will be used for uniform bands, WBATE, or LBATE, using \code{fitmethod = "joint"} unless the goal is a comparison with the earlier convention. Third, add \code{cluster} and \code{covs.eff} when the design calls for clustered inference or pre-treatment efficiency adjustment. Fourth, in fuzzy BD designs, inspect the FS BATEC before interpreting the fuzzy Wald curve, because weak or sign-changing first stages make the ratio unstable. Fifth, report pointwise intervals, uniform bands, WBATE, and LBATE for the main target, and use the side-specific effective sample sizes and bandwidth table to flag poorly supported boundary regions. Finally, compare the location-based analysis with distance-based estimates and with modest changes to the grid, bandwidth selector, and trimmed target set.

\subsection{Derivative Estimation}\label{sec: Location Derivative Estimation}

Location-based methods can also be used to estimate derivatives of the side-specific regression functions and their discontinuities at the boundary. For a multi-index $\bnu=(\nu_1,\nu_2)$ with $|\bnu|=\nu_1+\nu_2\leq p$, define the derivative BATEC
\begin{align*}
    \tau^{(\bnu)}(\bx)
    =
    \mu_1^{(\bnu)}(\bx)-\mu_0^{(\bnu)}(\bx),
    \qquad \bx\in\B,
\end{align*}
where $\mu_t^{(\bnu)}(\bx)$ denotes the partial derivative of $\mu_t(\bx)$ of order $\bnu$. These derivative targets describe how the discontinuity changes locally with the score coordinates; for example, $\bnu=(1,0)$ and $\bnu=(0,1)$ correspond to the two coordinate-specific slopes of the BATEC at the boundary.

The plug-in estimator uses the same local polynomial coefficients as the level estimator. Using the selector $\be_{\bnu}$ introduced above,
\begin{align*}
    \widehat{\tau}^{(\bnu)}(\bx)
    =
    \bnu!\,h^{-|\bnu|}
    \be_{\bnu}^{\top}
    \widehat{\bdelta}(\bx),
    \qquad \bx\in\B,
\end{align*}
where $\bnu!=\nu_1!\nu_2!$. Directional derivatives, including derivatives along the boundary, are obtained by applying the corresponding linear combination to the first partial derivatives.

In \pkg{rd2d}, derivative targets are requested with \code{deriv}; for example, \code{deriv = c(1, 0)} estimates the discontinuity in the first partial derivative with respect to the first score coordinate. Directional derivatives along user-supplied vectors, such as boundary tangent vectors, are requested with \code{tangvec}, a $J\times2$ matrix with one vector for each boundary evaluation point; when supplied, \code{tangvec} overrides \code{deriv}. In fuzzy BD designs, the derivative options also apply to the reduced-form and first-stage output tables. Ratios involving derivative discontinuities should be interpreted separately from the level fuzzy BATEC $\zeta(\bx)=\tau_Y(\bx)/\tau_W(\bx)$.

\subsection{Covariate Adjustment}

Predetermined covariates can improve precision in boundary discontinuity designs when they explain residual variation in the outcome. The implementation in \pkg{rd2d} follows the covariate adjustment logic developed for canonical RD designs by \citet{Calonico-Cattaneo-Farrell-Titiunik_2019_RESTAT}. Let $\bZ_i$ be a $d_Z$-vector of predetermined covariates. We use a tilde to denote covariate-adjusted quantities. The covariate-adjusted fit uses the same augmented local-polynomial basis as above and adds $\bZ_i^\top\blambda$ to the weighted least-squares regression:
\begin{align*}
    \left(\widetilde{\bdelta}(\bx),\widetilde{\blambda}(\bx)\right)
    =
    \argmin_{\bdelta\in\mathbb{R}^{2\mathfrak{p}_p},\,\blambda\in\mathbb{R}^{d_Z}}
    \frac{1}{n}\sum_{i=1}^n
    \big(Y_i-\br_{p,h}(\bX_i-\bx;T_i)^\top\bdelta-\bZ_i^\top\blambda\big)^2
    K_h(\bX_i-\bx).
\end{align*}
The covariate-adjusted BATEC estimator is $\widetilde{\tau}(\bx)=\be_{\mathbf{0}}^{\top}\widetilde{\bdelta}(\bx)$. Under the usual smoothness and no-manipulation conditions, this adjustment does not change the estimand; it can reduce variance when the covariates are predictive of the outcome.

In the software, covariates are supplied through \code{covs.eff}. The covariate coefficient is common across treatment sides, while the local polynomial basis remains side-specific. The implementation residualizes the covariates on the local polynomial basis and carries the resulting adjusted influence functions through pointwise inference, uniform bands, WBATE, LBATE, fuzzy ratio inference, and automatic bandwidth selection. Thus, when \code{h} is omitted, the same covariate adjustment used in estimation is also used by \code{rdbw2d()}. The options \code{covs.drop} and \code{covs.tol} control safeguards for rank-deficient covariate matrices after residualization.

\begin{lstlisting}[style=Rstyle]
# Z is an n by d matrix or data frame of pre-treatment covariates.
fit_cov <- rd2d(Y = Y, X = X, assignment = T, b = b,
                fuzzy = W, covs.eff = Z,
                fitmethod = "joint",
                params.cov = c("main", "itt", "fs"))
summary(fit_cov, output = "main", cbands = "main")

bw_cov <- rdbw2d(Y = Y, X = X, assignment = T, b = b,
                 fuzzy = W, covs.eff = Z,
                 bwparam = "main")
\end{lstlisting}

The boundary setting differs from the scalar RD setting because the local fit is repeated over points on $\B$ and because covariance across boundary points matters for uniform bands, WBATE, and LBATE. Thus, covariate adjustment must be carried through the same covariance calculations used for the unadjusted BATEC. The same command structure can also be used for falsification exercises by treating predetermined covariates as outcomes, but that diagnostic task is distinct from covariate-adjusted treatment-effect estimation. The empirical application below keeps the fuzzy BD analysis unadjusted to focus on the default bivariate and distance-based estimators.

\section{Distance-Based Methods}\label{sec: Distance-Based Methods}

Distance-based methods replace the bivariate score $\bX_i$ by a scalar signed distance score constructed separately for each boundary point. For $\bx=(x_1,x_2)^\top\in\B$, let
\begin{align*}
    D_i(\bx)
    =
    \big(\Indicator(\bX_i\in\A_1)-\Indicator(\bX_i\in\A_0)\big)\d(\bX_i,\bx),
\end{align*}
where $\d(\cdot,\cdot)$ is a distance function such as the Euclidean distance $\d(\bX_i,\bx)=\|\bX_i-\bx\|$. Thus, $D_i(\bx)\ge0$ identifies observations on the treated side of the boundary and $D_i(\bx)<0$ identifies observations on the control side. Conditional on $\bx$, the problem resembles a univariate RD design with cutoff zero, while varying $\bx$ over $\B$ recovers a curve of local effects. In the Opportunity Zones example, each column of the distance matrix is the signed distance from a tract's poverty-income score to one of the 31 boundary points shown in Figure~\ref{fig:empapp-scatter}.

This construction is attractive when the full bivariate score is unavailable, difficult to release, or inconvenient to smooth directly. It also gives an intuitive set of RD plots and diagnostics for each boundary point. The cost is that distance-based methods use less location information than the location-based procedures in Section~\ref{sec: Location-Based Methods}; as a result, their interpretation depends more explicitly on the geometry of $\B$ and on the chosen distance function.

Distance-based identification requires an additional step because the scalar score averages potential outcomes over level sets of the distance function. For $\bx\in\B$, define
\begin{align*}
    \theta_{t,\bx}(r)
    = \E\big[Y_i \mid D_i(\bx) = r\big]
    = \E\big[Y_i(t) \mid \d(\bX_i,\bx) = |r|,\bX_i\in\A_t\big],
    \qquad r\in\I_t,
\end{align*}
where $\I_0=(-\infty,0)$ and $\I_1=[0,\infty)$. The induced signed-distance treatment-effect parameter is
\begin{align*}
    \vartheta(\bx)
    =
    \theta_{1,\bx}(0)-\theta_{0,\bx}(0),
    \qquad \bx\in\B.
\end{align*}
\citet[Theorem~1]{Cattaneo-Titiunik-Yu_2026_JOE} gives conditions on the data-generating process, the assignment boundary, the kernel, and the distance function under which
\begin{align*}
    \vartheta(\bx) = \tau(\bx),
    \qquad \bx\in\B.
\end{align*}
The result is obtained by representing the induced conditional expectations as integrals over shrinking submanifolds around $\bx$; without those restrictions on the distribution, the boundary geometry, and the distance function, the distance-based estimand need not coincide with the BATEC.

The same logic identifies aggregate and fuzzy targets. Since $\vartheta(\bx)=\tau(\bx)$ pointwise along $\B$ under the distance-based identification conditions, WBATE and LBATE can be represented by applying the same integral and supremum functionals to $\vartheta(\bx)$. In fuzzy BD designs, applying the signed-distance construction to $A\in\{Y,W\}$ gives
\begin{align*}
    \vartheta_A(\bx)
    =
    \theta_{A,1,\bx}(0)-\theta_{A,0,\bx}(0),
    \qquad
    \xi(\bx)=\frac{\vartheta_Y(\bx)}{\vartheta_W(\bx)}.
\end{align*}
Under the fuzzy identification conditions in \citet[Section~SA-6]{Cattaneo-Titiunik-Yu_2026_JOE}, $\vartheta_A(\bx)=\tau_A(\bx)$ for $A\in\{Y,W\}$, and if the first stage is nonzero then $\xi(\bx)=\zeta(\bx)$. Thus, aggregating $\xi(\bx)$ or taking its supremum gives distance-based representations of the causal fuzzy WBATE and fuzzy LBATE.

\subsection{BATEC Estimation and Inference}\label{sec: Distance BATEC}

The distance-based local polynomial estimator is computed separately for every evaluation point. In sharp BD designs, the analysis replaces \code{X} with an $n$ by $J$ matrix of signed distance scores, one column per boundary point. For Opportunity Zones, the $j$th column of \code{D} is the signed Euclidean distance from each tract's poverty-income score to $\bb_j$, with positive values for eligible tracts:
\begin{lstlisting}[style=Rstyle]
# D is an n by J signed-distance matrix, one column per boundary point.
fit_dist <- rd2d.distance(Y = Y, distance = D, b = b,
                          cbands = TRUE)
summary(fit_dist, output = "main", cbands = "main")
\end{lstlisting}

The point estimator is
\begin{align*}
    \widehat{\vartheta}(\bx)
    = \be_0^{\top} \widehat{\bgamma}_1(\bx)
      - \be_0^{\top} \widehat{\bgamma}_0(\bx),
    \qquad \bx \in \B,
\end{align*}
where, for $t \in \{0,1\}$,
\begin{align*}
    \widehat{\bgamma}_{t}(\bx)
    = \argmin_{\bgamma \in \mathbb{R}^{p+1}}
      \frac{1}{n}\sum_{i=1}^n
      \big(Y_i-\br_p(D_i(\bx)/h)^\top\bgamma\big)^2
      K_h(D_i(\bx))\Indicator(D_i(\bx)\in\I_t),
\end{align*}
with $\be_0$ the intercept selector, $\br_p(u)=(1,u,u^2,\ldots,u^p)^\top$ the usual univariate polynomial basis, $K_h(u)=K(u/h)/h^2$, and $\I_0=(-\infty,0)$ and $\I_1=[0,\infty)$. The normalization by $h^2$ reflects the dimension of the original bivariate score; it does not change the weighted least squares solution because it multiplies all weights by the same positive constant.

\citet{Cattaneo-Titiunik-Yu_2026_JOE} establishes pointwise and uniform consistency, distributional approximations, and confidence intervals and bands for $\widehat{\vartheta}(\bx)$. The main practical distinction relative to location-based methods is the bias behavior near nonsmooth regions of the boundary. If $\B$ is sufficiently smooth, the distance-induced regression functions have the usual local-polynomial approximation properties, and robust bias correction proceeds with $q=p+1$ by default. If $\B$ has a kink or another local irregularity, the induced univariate regression can be only Lipschitz even when the underlying bivariate regression functions are smoother. Then the leading bias is of order $h$ uniformly near the irregularity, and increasing $p$ does not remove it.

\subsection{WBATE and LBATE}\label{sec: Distance Aggregate Effects}

The causal WBATE and LBATE remain the functionals of the BATEC defined in Section~\ref{sec: Setup}. Under the distance-based identification conditions, $\tau(\bx)=\vartheta(\bx)$ for $\bx\in\B$, so the same targets can be written using the signed-distance estimand. With normalized weights $\int_{\B}w(\bx)\dif\bx=1$,
\begin{align*}
    \tau_{\mathtt{WBATE}}
    =
    \int_{\B}\tau(\bx)w(\bx)\dif\bx
    =
    \int_{\B}\vartheta(\bx)w(\bx)\dif\bx,
    \qquad
    \widehat{\vartheta}_{\mathtt{WBATE}}
    =
    \int_{\B}\widehat{\vartheta}(\bx)w(\bx)\dif\bx.
\end{align*}
On the evaluation grid, $\widehat{\vartheta}_{\mathtt{WBATE}}$ is implemented as a weighted average of the distance-based BATEC estimates. The covariance is the corresponding double integral of the distance-based covariance kernel,
\begin{align*}
    \widehat{\Xi}_{\mathtt{WBATE}}
    =
    \int_{\B}\int_{\B}
    \widehat{\Xi}_{\bx_1,\bx_2}
    w(\bx_1)w(\bx_2)\dif\bx_1\dif\bx_2,
\end{align*}
or its grid analogue. As shown in \citet[Section~SA-4]{Cattaneo-Titiunik-Yu_2026_JOE}, the WBATE has effective variance order $(nh)^{-1}$ under local boundary-measure regularity, even though each pointwise distance-based fit has the bivariate variance order $(nh^2)^{-1}$ when $d=2$.

Similarly, under distance-based identification, the causal LBATE can be written as
\begin{align*}
    \tau_{\mathtt{LBATE}}
    =
    \sup_{\bx\in\B}\tau(\bx)
    =
    \sup_{\bx\in\B}\vartheta(\bx),
    \qquad
    \widehat{\vartheta}_{\mathtt{LBATE}}
    =
    \sup_{\bx\in\B}\widehat{\vartheta}(\bx).
\end{align*}
Its confidence interval is obtained by projecting the uniform confidence band for $\vartheta(\bx)$, as in \citet[Section~SA-5]{Cattaneo-Titiunik-Yu_2026_JOE}. The resulting interval is conservative in general because it inherits simultaneous coverage for the full distance-based BATEC.

In the software, both summaries are produced by the \code{summary()} method once the relevant covariance matrix has been stored:
\begin{lstlisting}[style=Rstyle]
w <- rep(1, nrow(b))
summary(fit_dist, output = "main", cbands = "main",
        WBATE = w, LBATE = TRUE)
\end{lstlisting}
Under the distance-based identification conditions, these are plug-in estimators of $\tau_{\mathtt{WBATE}}$ and $\tau_{\mathtt{LBATE}}$.

\subsection{Imperfect Compliance}\label{sec: Distance Fuzzy Methods}

Distance-based methods extend to fuzzy BD designs by applying the same signed-distance local polynomial procedure to the outcome and to treatment receipt. For $A\in\{Y,W\}$, let
\begin{align*}
    \widehat{\vartheta}_A(\bx)
    =
    \be_0^\top\widehat{\bgamma}_{A,1}(\bx)
    -
    \be_0^\top\widehat{\bgamma}_{A,0}(\bx),
\end{align*}
where $\widehat{\bgamma}_{A,t}(\bx)$ is the local polynomial fit above with $A_i$ replacing $Y_i$. The distance-based fuzzy BATEC estimator is the Wald ratio
\begin{align*}
    \widehat{\xi}(\bx)
    = \frac{\widehat{\vartheta}_{Y}(\bx)}
           {\widehat{\vartheta}_{W}(\bx)},
    \qquad \bx\in\B.
\end{align*}
The reduced-form output estimates $\vartheta_Y(\bx)$, the first-stage output estimates $\vartheta_W(\bx)$, and the main fuzzy output estimates $\xi(\bx)$. Under the identification conditions summarized above, these correspond to $\tau_Y(\bx)$, $\tau_W(\bx)$, and the fuzzy BATEC $\zeta(\bx)$.

Inference is based on the ratio linearization in \citet[Section~SA-6]{Cattaneo-Titiunik-Yu_2026_JOE}. With
\begin{align*}
    \mathfrak{w}(\bx)
    =
    \left[
    \frac{1}{\vartheta_W(\bx)},
    -\frac{\vartheta_Y(\bx)}{\vartheta_W(\bx)^2}
    \right]^\top,
\end{align*}
the first-order component of $\widehat{\xi}(\bx)-\xi(\bx)$ is the linear combination of the reduced-form and first-stage errors indexed by $\mathfrak{w}(\bx)$. Thus, the fuzzy covariance kernel is obtained by sandwiching the joint covariance kernel of $(\widehat{\vartheta}_Y(\bx),\widehat{\vartheta}_W(\bx))^\top$ with this gradient vector. This delta-method covariance is used for pointwise intervals, uniform confidence bands, fuzzy WBATE inference, and fuzzy LBATE inference.

With the same weight normalization, the causal fuzzy WBATE and fuzzy LBATE remain the functionals of $\zeta(\bx)$ defined in Section~\ref{sec: Setup},
\begin{align*}
    \zeta_{\mathtt{WBATE}}
    =
    \int_{\B}\zeta(\bx)w(\bx)\dif\bx,
    \qquad
    \zeta_{\mathtt{LBATE}}
    =
    \sup_{\bx\in\B}\zeta(\bx).
\end{align*}
The corresponding distance-based plug-in estimators replace $\zeta(\bx)$ by $\widehat{\xi}(\bx)$:
\begin{align*}
    \widehat{\xi}_{\mathtt{WBATE}}
    =
    \int_{\B}\widehat{\xi}(\bx)w(\bx)\dif\bx,
    \qquad
    \widehat{\xi}_{\mathtt{LBATE}}
    =
    \sup_{\bx\in\B}\widehat{\xi}(\bx).
\end{align*}
Under fuzzy distance-based identification, $\xi(\bx)=\zeta(\bx)$, so these are plug-in estimators of $\zeta_{\mathtt{WBATE}}$ and $\zeta_{\mathtt{LBATE}}$. In the Opportunity Zones application, the FS curve measures how eligibility changes the probability of designation, the ITT curve measures how eligibility changes early workplace-job growth, and the fuzzy curve rescales the latter by the former.

In the software, supplying \code{fuzzy} changes \code{main} to the fuzzy Wald effect, while \code{itt} and \code{fs} report the reduced-form and first-stage BATECs. The option \code{bwparam = "main"} selects automatic bandwidths for the linearized fuzzy ratio, while \code{bwparam = "itt"} targets the reduced-form outcome.
\begin{lstlisting}[style=Rstyle]
fit_dist_fuzzy <- rd2d.distance(
    Y = Y, distance = D, b = b, fuzzy = W,
    params.cov = c("main", "itt", "fs"))

summary(fit_dist_fuzzy, output = "itt",  cbands = "itt")
summary(fit_dist_fuzzy, output = "fs",   cbands = "fs")
summary(fit_dist_fuzzy, output = "main", cbands = "main",
        WBATE = w, LBATE = TRUE)
\end{lstlisting}

\subsection{Implementation Details}\label{sec: Distance Implementation Details}

The distance-based commands share most options with the location-based commands. Bandwidth selection uses \code{rdbw2d.distance()}; if \code{h} is omitted, \code{rd2d.distance()} calls this selector internally. A scalar \code{h} applies the same bandwidth to all boundary points and both sides, while a $J$ by $2$ matrix supplies boundary-specific control and treated bandwidths. Unless overridden, \code{rd2d.distance()} uses a triangular kernel, $p=1$, $q=p+1$ under the smooth-boundary default, \code{vce = "hc1"}, \code{fitmethod = "joint"}, \code{masspoints = "check"}, \code{level = 95}, and automatic bandwidth selection with \code{bwselect = "mserd"}. The available \code{bwselect} choices are \code{mserd}, \code{cerrd}, \code{imserd}, \code{icerrd}, \code{msetwo}, \code{certwo}, \code{imsetwo}, and \code{icertwo}; the default regularization constants are \code{scaleregul = 1}, \code{cqt = 0.5}, and \code{bwcheck = 50 + p + 1} for \code{rd2d.distance()}.

Cluster-robust inference and covariate efficiency adjustment use the same option names and interpretation as in the location-based commands. The default \code{fitmethod = "joint"} aggregates cluster scores across both signs of the distance score, which is the appropriate convention when clusters can appear on both sides of a local distance cutoff. The option \code{covs.eff} supplies predetermined covariates for the common-coefficient efficiency adjustment and is also passed to \code{rdbw2d.distance()} when automatic bandwidths are selected.

\begin{lstlisting}[style=Rstyle]
fit_dist_cov <- rd2d.distance(
    Y = Y, distance = D, b = b, fuzzy = W,
    cluster = cluster_id,
    covs.eff = Z,
    fitmethod = "joint",
    params.cov = c("main", "itt", "fs"))
\end{lstlisting}

The package exposes the boundary-geometry distinction in \citet{Cattaneo-Titiunik-Yu_2026_JOE} through two options. The default setting treats the boundary locally as smooth and uses the usual robust bias-corrected construction. When kinks may be present but their locations are not being used, \code{kink.unknown} can shrink the automatic bandwidths to the rates justified for unknown irregularities and set the default bias-correction order to $q=p$. When kink locations are known on the evaluation grid, \code{kink.position} accepts either a logical vector or integer indices of the kink boundary points, and automatic bandwidth selection adapts to proximity to those locations:
\begin{lstlisting}[style=Rstyle]
fit_kink <- rd2d.distance(Y = Y, distance = D, b = b,
                          kink.position = 16)
summary(fit_kink, output = "main", cbands = "main")

fit_ukink <- rd2d.distance(Y = Y, distance = D, b = b,
                           kink.unknown = c(TRUE, TRUE))
summary(fit_ukink, output = "main", cbands = "main")
\end{lstlisting}

Additional output can be requested when needed. In sharp BD designs, \code{params.other} can request \code{main.0} and \code{main.1}, the two side-specific fits. In fuzzy BD designs, \code{params.other} can request \code{itt.0}, \code{itt.1}, \code{fs.0}, and \code{fs.1}; \code{params.cov} controls which covariance matrices are stored for uniform bands and aggregate inference. The printed and returned \code{summary()} tables mirror the location-based output: point estimates, RBC standard errors, test statistics, $p$-values, pointwise intervals, optional uniform-band columns, bandwidths, and effective sample sizes are reported at each boundary point, with optional WBATE and LBATE rows appended. Standard error options include \code{vce = "hc0"}, \code{"hc1"}, \code{"hc2"}, or \code{"hc3"}. Finally, \code{masspoints} controls checking and adjustment for repeated signed-distance values, which can matter when locations or distances are discretized. These options make distance-based analysis useful both as a standalone strategy and as a robustness comparison to the richer location-based methods.

\section{Empirical Results}\label{sec: Empirical Application}

We now return to the Opportunity Zones running example introduced in Section~\ref{sec: Running Example}. The empirical analysis reports the three objects in a fuzzy BD design: the reduced-form BATEC for early workplace-job growth, the first-stage BATEC for Qualified Opportunity Zone designation, and the fuzzy BATEC obtained by dividing the reduced form by the first stage. We report the fuzzy curve as a Wald estimand; a complier-average causal interpretation would require the additional local exclusion, monotonicity, and nonzero-first-stage conditions discussed in Section~\ref{sec: Setup}. For each object, we compare the location-based fit, which uses the bivariate poverty-income score directly, with the distance-based fit, which uses the signed Euclidean distance to each evaluation point on the boundary. Both fits use the package's automatic default choices, including \code{fitmethod = "joint"}; the only additional request is to store covariance matrices for uniform bands and aggregate summaries. We do not use \code{covs.eff} in this illustration, so that the reported results correspond to the unadjusted fuzzy BD estimators.

\begin{figure}[p]
    \centering
    \begin{subfigure}[b]{0.48\textwidth}
        \centering
        \includegraphics[width=\linewidth]{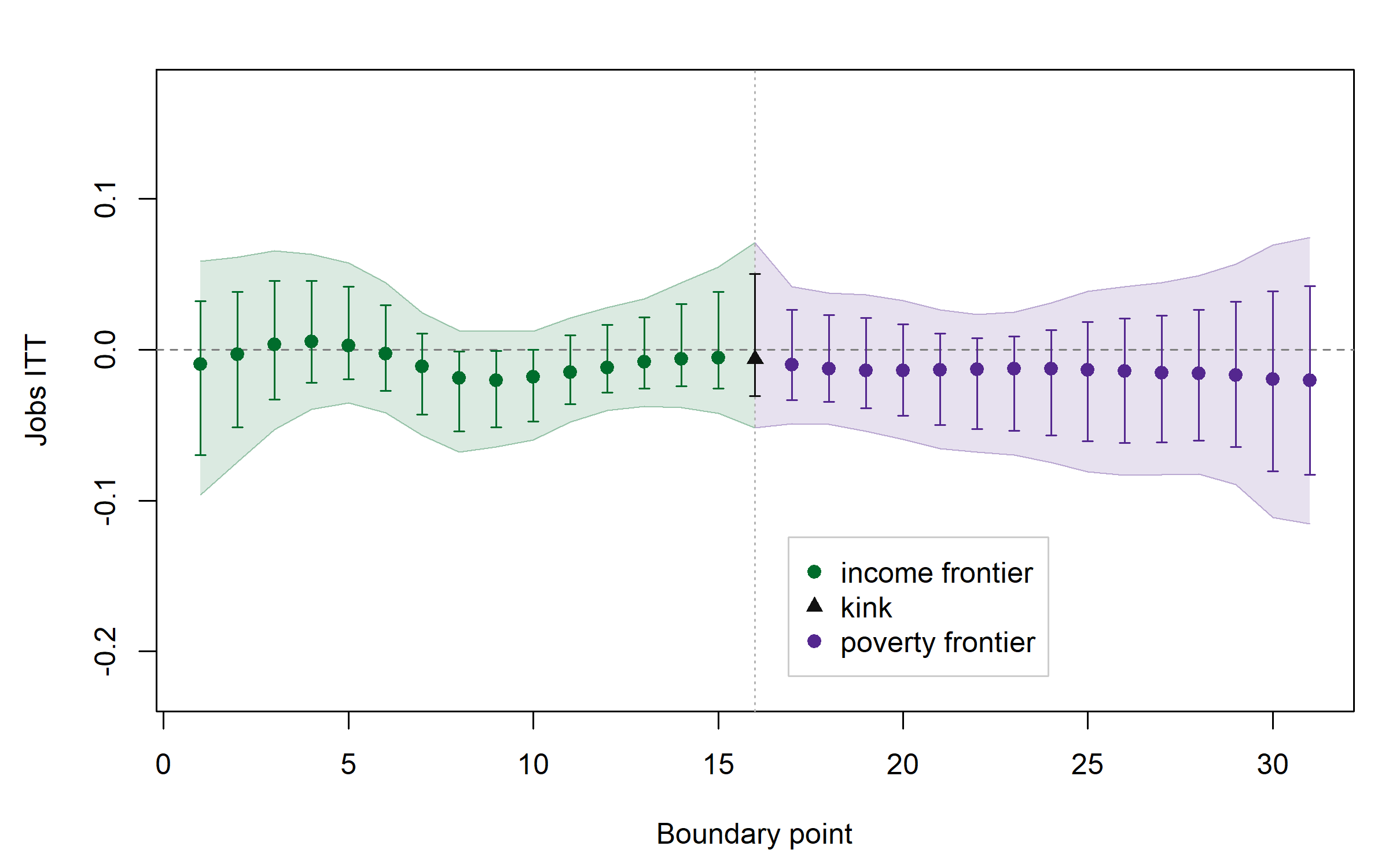}
        \caption{Location-Based ITT}
    \end{subfigure}
    \hfill
    \begin{subfigure}[b]{0.48\textwidth}
        \centering
        \includegraphics[width=\linewidth]{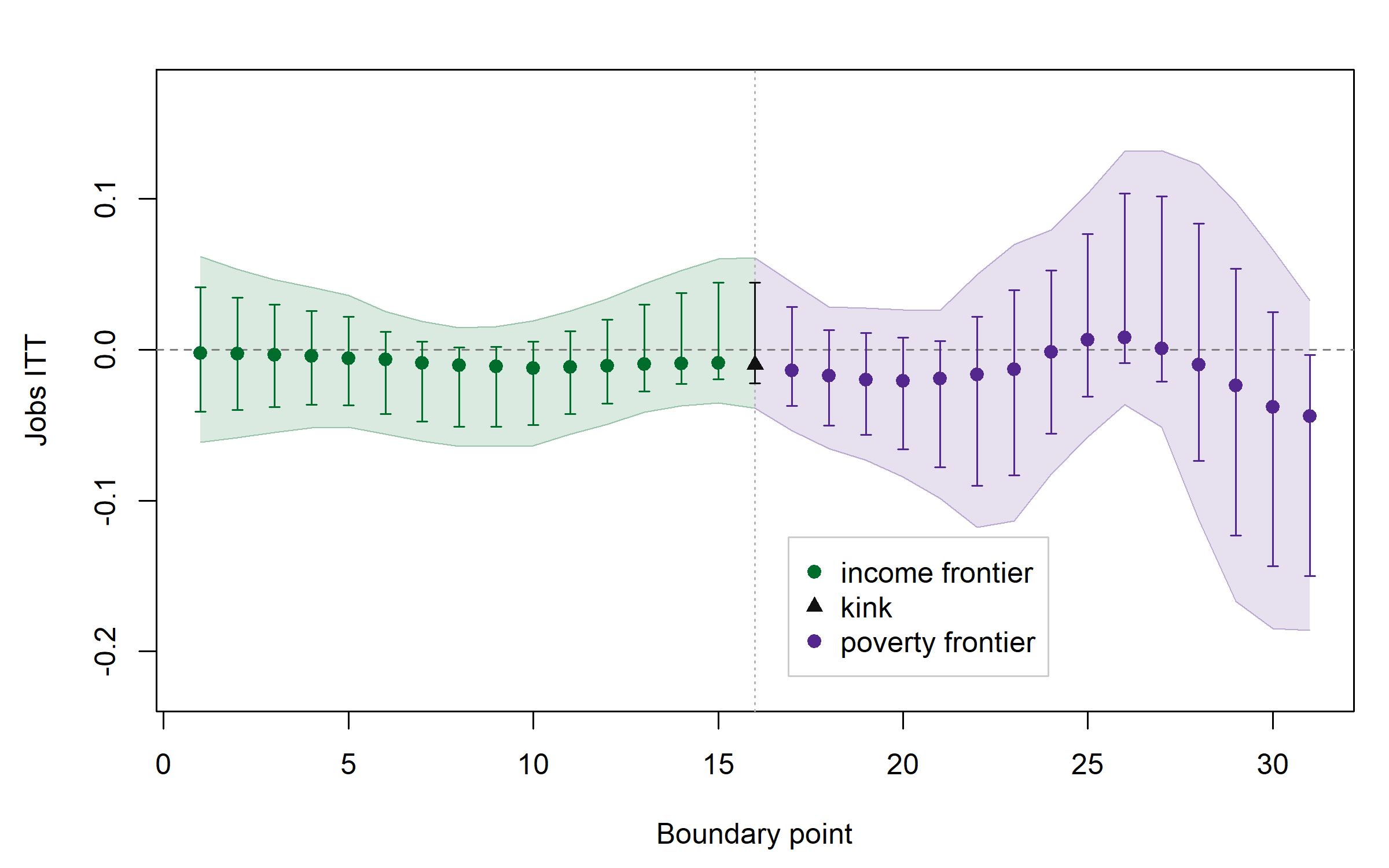}
        \caption{Distance-Based ITT}
    \end{subfigure}

    \vspace{0.25cm}
    \begin{subfigure}[b]{0.48\textwidth}
        \centering
        \includegraphics[width=\linewidth]{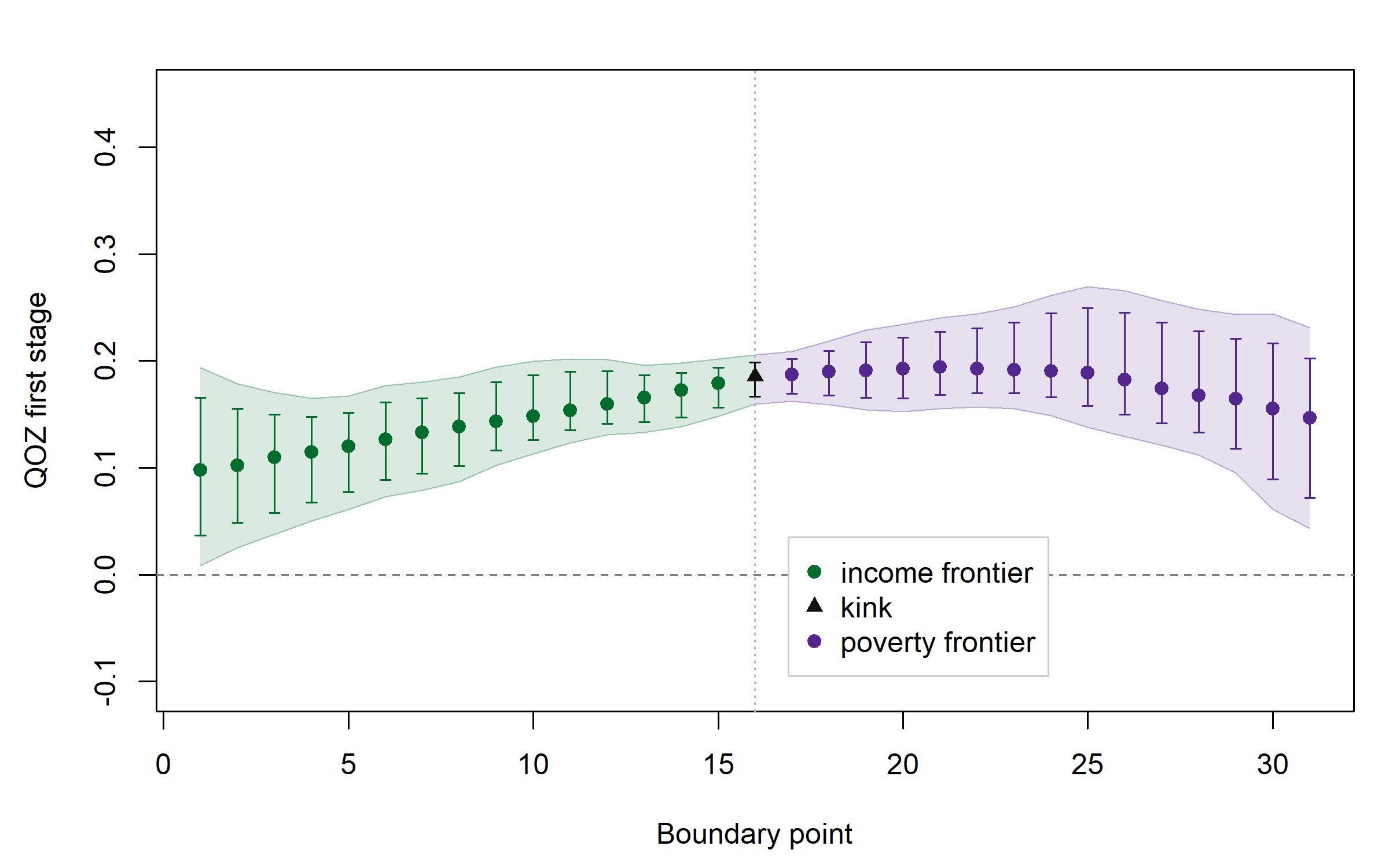}
        \caption{Location-Based First Stage}
    \end{subfigure}
    \hfill
    \begin{subfigure}[b]{0.48\textwidth}
        \centering
        \includegraphics[width=\linewidth]{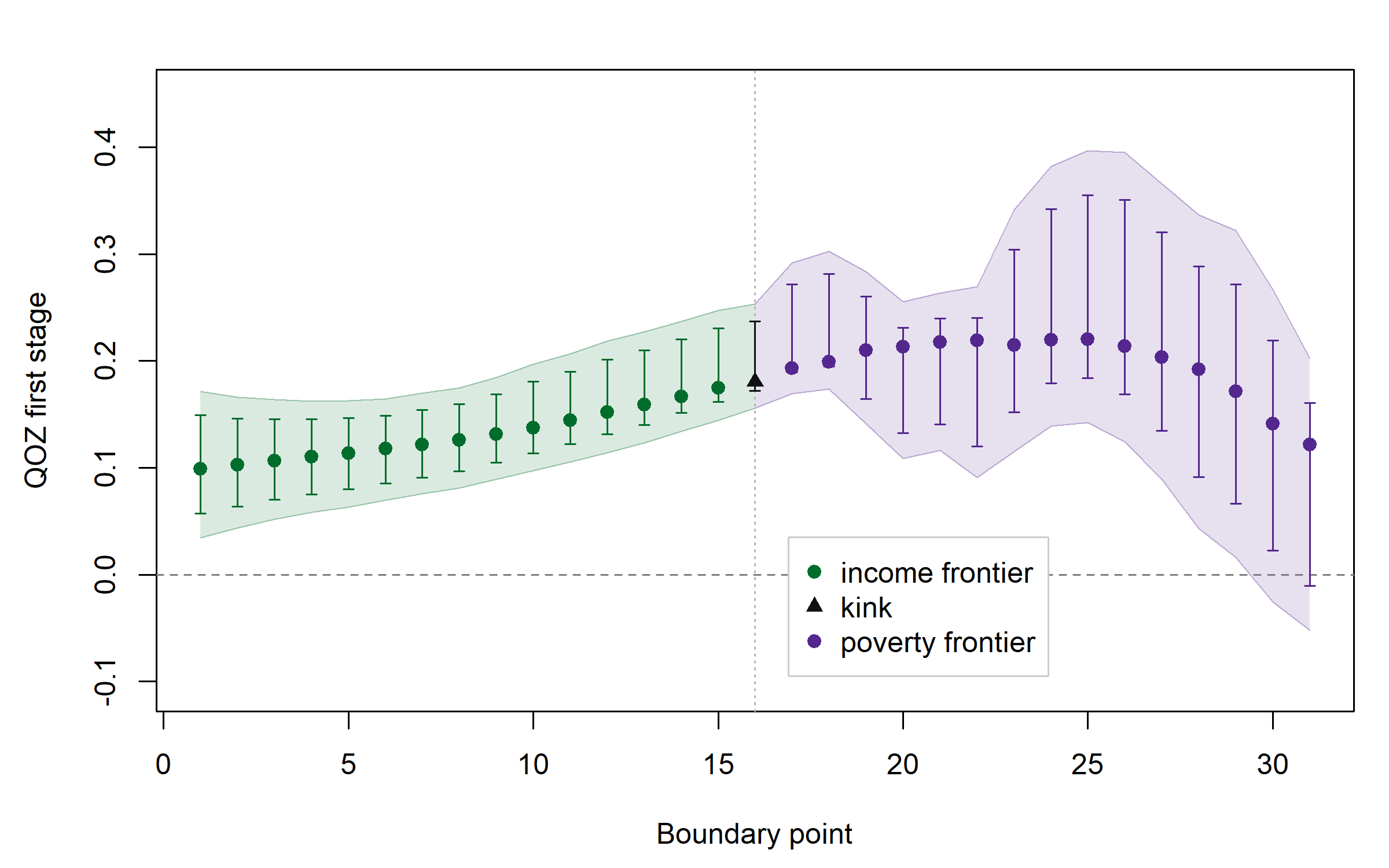}
        \caption{Distance-Based First Stage}
    \end{subfigure}

    \vspace{0.25cm}
    \begin{subfigure}[b]{0.48\textwidth}
        \centering
        \includegraphics[width=\linewidth]{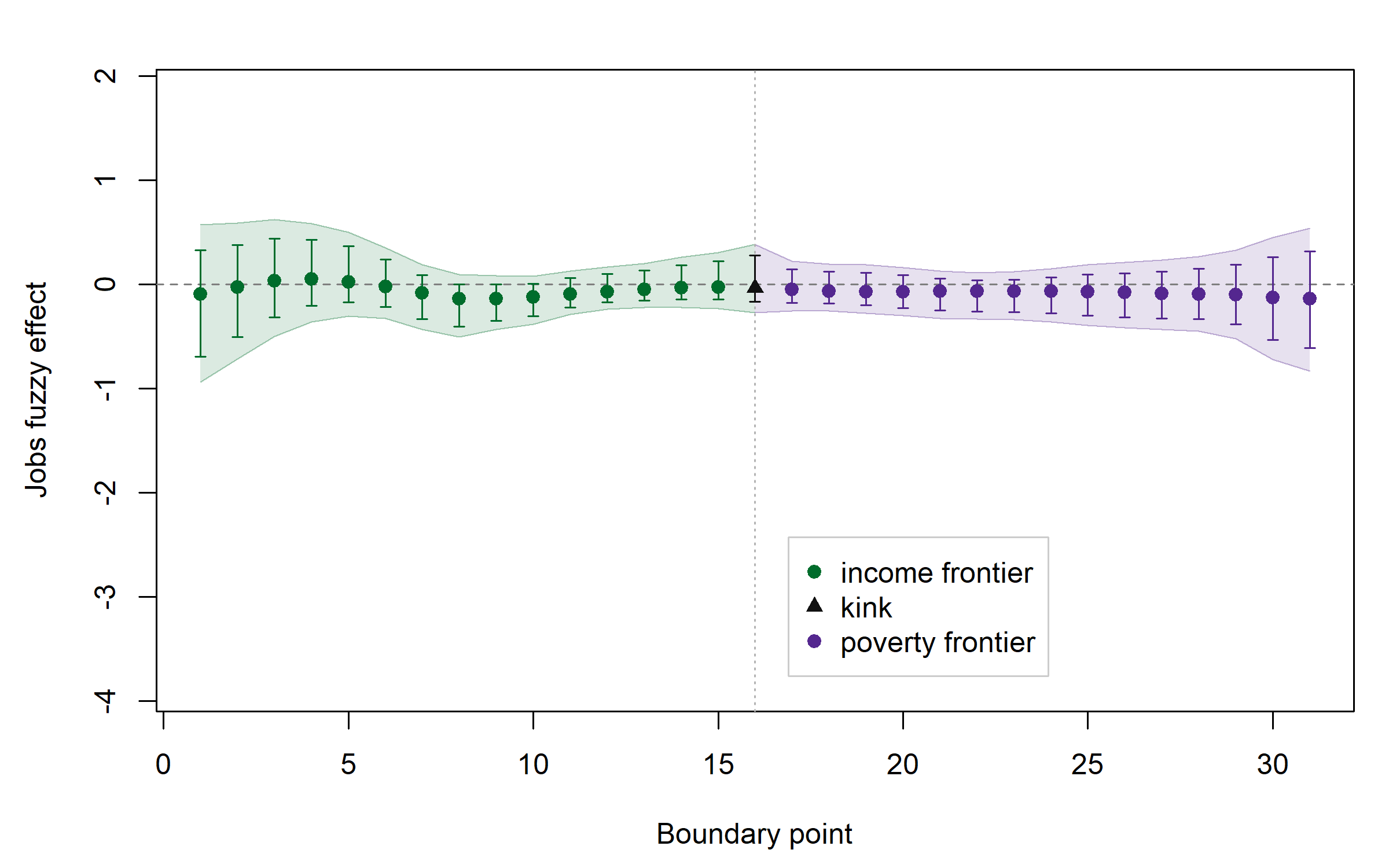}
        \caption{Location-Based Fuzzy BATEC}
    \end{subfigure}
    \hfill
    \begin{subfigure}[b]{0.48\textwidth}
        \centering
        \includegraphics[width=\linewidth]{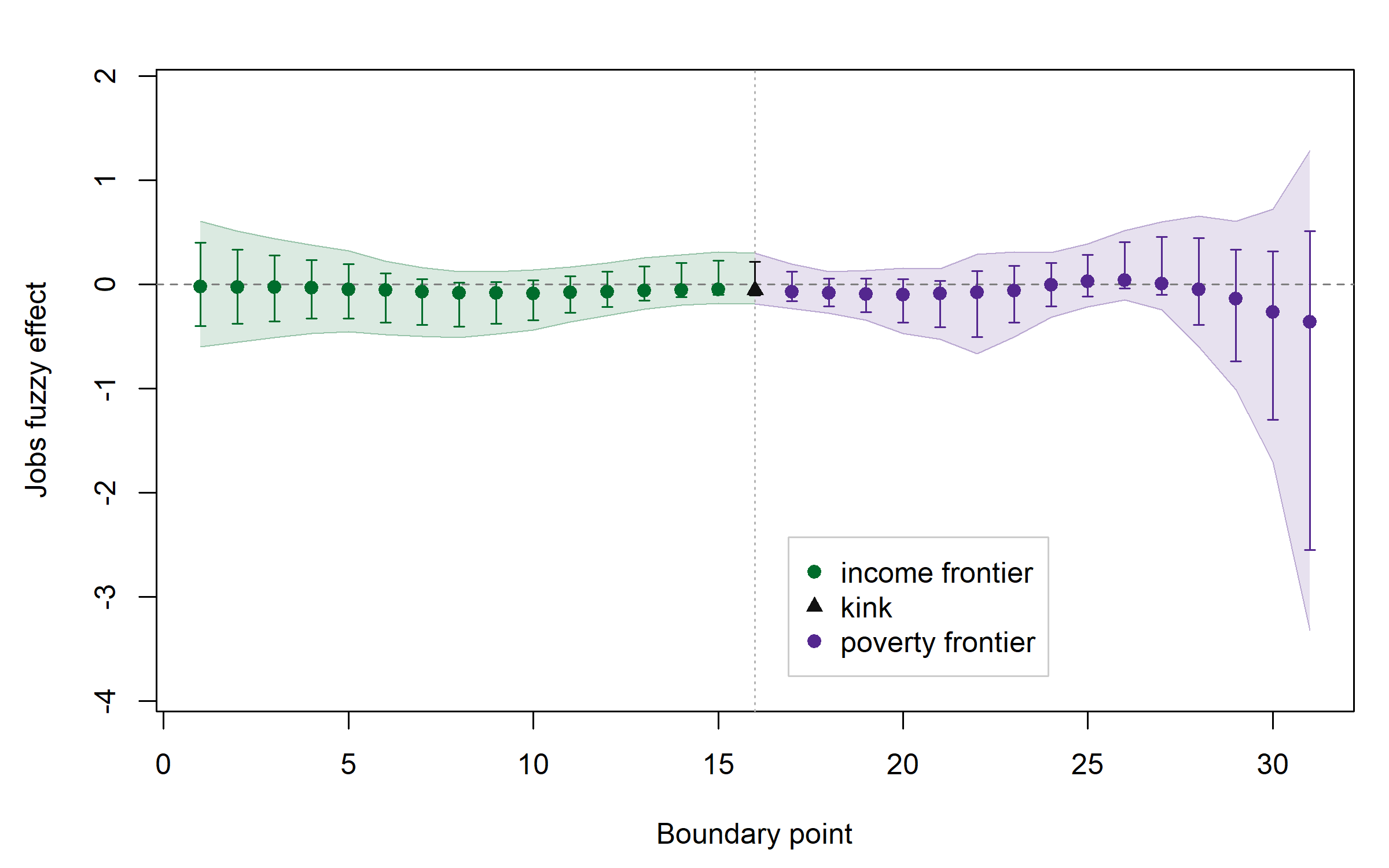}
        \caption{Distance-Based Fuzzy BATEC}
    \end{subfigure}
    \caption{Opportunity Zones Empirical Results. Rows report the reduced-form BATEC, first-stage BATEC, and fuzzy BATEC; columns compare location-based and distance-based methods. Bands are uniform confidence bands over the 31-point boundary grid shown in Figure~\ref{fig:empapp-scatter}.}
    \label{fig:empapp-results}
\end{figure}

Figure~\ref{fig:empapp-results} gives the main visual comparison. The rows correspond to the reduced form, first stage, and fuzzy effect; the columns correspond to the location-based and distance-based methods. Both methods find a strong first stage along most of the boundary: eligibility sharply increases the probability of Qualified Opportunity Zone designation. In contrast, the reduced-form BATEC for early workplace-job growth is small relative to its uncertainty, and the fuzzy BATEC is correspondingly imprecise. The distance-based first stage is slightly less uniform near some boundary locations, which is consistent with the sensitivity to boundary geometry discussed in Section~\ref{sec: Distance-Based Methods}.

Tables~\ref{tab:empapp-output-itt}--\ref{tab:empapp-output-fuzzy} report selected numerical output underlying Figure~\ref{fig:empapp-results}. To conserve space, each table reports every fifth boundary point, the kink, the two endpoints, and the WBATE and LBATE rows generated by \code{summary()}. The full tables are generated by the replication script and written to the \code{inputs} folder.

\begin{table}[p]
    \centering
    \caption{Opportunity Zones Reduced-Form BATEC}
    \label{tab:empapp-output-itt}
    \resizebox{\textwidth}{!}{\begin{tabular}{lccc ccc}
  \toprule\toprule
   & \multicolumn{3}{c}{Location-Based} & \multicolumn{3}{c}{Distance-Based}\\
  \cmidrule(lr){2-4}\cmidrule(lr){5-7}
   & Estimate & $p$-value & 95\% CI & Estimate & $p$-value & 95\% CI\\
  \midrule
   $\bb_{1}$ & -0.019 & 0.471 & (-0.070, 0.032) & 0.000 & 0.996 & (-0.041, 0.041)\\
   $\bb_{5}$ & 0.011 & 0.475 & (-0.019, 0.042) & -0.008 & 0.611 & (-0.037, 0.022)\\
   $\bb_{10}$ & -0.024 & 0.050 & (-0.048, -0.000) & -0.022 & 0.115 & (-0.050, 0.005)\\
   $\bb_{15}$ & 0.006 & 0.696 & (-0.026, 0.038) & 0.013 & 0.442 & (-0.019, 0.045)\\
   $\bb_{16}$ & 0.010 & 0.637 & (-0.031, 0.050) & 0.011 & 0.510 & (-0.022, 0.045)\\
   $\bb_{20}$ & -0.014 & 0.382 & (-0.044, 0.017) & -0.029 & 0.125 & (-0.066, 0.008)\\
   $\bb_{25}$ & -0.021 & 0.294 & (-0.061, 0.018) & 0.023 & 0.406 & (-0.031, 0.077)\\
   $\bb_{30}$ & -0.021 & 0.492 & (-0.081, 0.039) & -0.059 & 0.167 & (-0.143, 0.025)\\
   $\bb_{31}$ & -0.020 & 0.524 & (-0.083, 0.042) & -0.077 & 0.040 & (-0.150, -0.004)\\
  \midrule
   $\mathtt{WBATE}$ & -0.011 & 0.227 & (-0.029, 0.007) & -0.011 & 0.185 & (-0.027, 0.005)\\
  \midrule
   $\mathtt{LBATE}$ & 0.012 &  & (-0.035, 0.074) & 0.047 &  & (-0.037, 0.135)\\
  \bottomrule\bottomrule
\end{tabular}
}
    {\footnotesize \textit{Notes}. The outcome is the 2017--2019 change in log workplace jobs. The WBATE estimator uses equal weights over the boundary grid points.}
\end{table}

\begin{table}[p]
    \centering
    \caption{Opportunity Zones First-Stage BATEC}
    \label{tab:empapp-output-fs}
    \resizebox{\textwidth}{!}{\begin{tabular}{lccc ccc}
  \toprule\toprule
   & \multicolumn{3}{c}{Location-Based} & \multicolumn{3}{c}{Distance-Based}\\
  \cmidrule(lr){2-4}\cmidrule(lr){5-7}
   & Estimate & $p$-value & 95\% CI & Estimate & $p$-value & 95\% CI\\
  \midrule
   $\bb_{1}$ & 0.101 & 0.002 & (0.037, 0.165) & 0.103 & 0.000 & (0.057, 0.149)\\
   $\bb_{5}$ & 0.114 & 0.000 & (0.077, 0.151) & 0.113 & 0.000 & (0.080, 0.147)\\
   $\bb_{10}$ & 0.156 & 0.000 & (0.126, 0.187) & 0.147 & 0.000 & (0.114, 0.181)\\
   $\bb_{15}$ & 0.175 & 0.000 & (0.156, 0.193) & 0.196 & 0.000 & (0.162, 0.231)\\
   $\bb_{16}$ & 0.183 & 0.000 & (0.167, 0.199) & 0.204 & 0.000 & (0.172, 0.237)\\
   $\bb_{20}$ & 0.194 & 0.000 & (0.165, 0.222) & 0.182 & 0.000 & (0.133, 0.231)\\
   $\bb_{25}$ & 0.204 & 0.000 & (0.158, 0.250) & 0.269 & 0.000 & (0.184, 0.355)\\
   $\bb_{30}$ & 0.153 & 0.000 & (0.089, 0.216) & 0.121 & 0.016 & (0.022, 0.219)\\
   $\bb_{31}$ & 0.137 & 0.000 & (0.072, 0.203) & 0.075 & 0.085 & (-0.010, 0.161)\\
  \midrule
   $\mathtt{WBATE}$ & 0.163 & 0.000 & (0.143, 0.182) & 0.171 & 0.000 & (0.152, 0.191)\\
  \midrule
   $\mathtt{LBATE}$ & 0.205 &  & (0.162, 0.271) & 0.269 &  & (0.174, 0.397)\\
  \bottomrule\bottomrule
\end{tabular}
}
    {\footnotesize \textit{Notes}. The outcome is Qualified Opportunity Zone designation. The WBATE estimator uses equal weights over the boundary grid points.}
\end{table}

\begin{table}[p]
    \centering
    \caption{Opportunity Zones Fuzzy BATEC}
    \label{tab:empapp-output-fuzzy}
    \resizebox{\textwidth}{!}{\begin{tabular}{lccc ccc}
  \toprule\toprule
   & \multicolumn{3}{c}{Location-Based} & \multicolumn{3}{c}{Distance-Based}\\
  \cmidrule(lr){2-4}\cmidrule(lr){5-7}
   & Estimate & $p$-value & 95\% CI & Estimate & $p$-value & 95\% CI\\
  \midrule
   $\bb_{1}$ & -0.186 & 0.475 & (-0.696, 0.324) & 0.001 & 0.996 & (-0.399, 0.401)\\
   $\bb_{5}$ & 0.098 & 0.480 & (-0.173, 0.369) & -0.067 & 0.612 & (-0.326, 0.192)\\
   $\bb_{10}$ & -0.153 & 0.055 & (-0.309, 0.003) & -0.152 & 0.121 & (-0.344, 0.040)\\
   $\bb_{15}$ & 0.036 & 0.696 & (-0.146, 0.219) & 0.064 & 0.443 & (-0.100, 0.228)\\
   $\bb_{16}$ & 0.053 & 0.637 & (-0.168, 0.275) & 0.055 & 0.511 & (-0.109, 0.218)\\
   $\bb_{20}$ & -0.070 & 0.383 & (-0.227, 0.087) & -0.160 & 0.131 & (-0.367, 0.048)\\
   $\bb_{25}$ & -0.104 & 0.299 & (-0.299, 0.092) & 0.085 & 0.407 & (-0.116, 0.285)\\
   $\bb_{30}$ & -0.137 & 0.498 & (-0.533, 0.259) & -0.492 & 0.233 & (-1.300, 0.317)\\
   $\bb_{31}$ & -0.149 & 0.530 & (-0.612, 0.315) & -1.020 & 0.192 & (-2.552, 0.511)\\
  \midrule
   $\mathtt{WBATE}$ & -0.066 & 0.263 & (-0.180, 0.049) & -0.099 & 0.124 & (-0.226, 0.027)\\
  \midrule
   $\mathtt{LBATE}$ & 0.110 &  & (-0.217, 0.609) & 0.183 &  & (-0.150, 1.282)\\
  \bottomrule\bottomrule
\end{tabular}
}
    {\footnotesize \textit{Notes}. The fuzzy BATEC divides the reduced-form BATEC by the first-stage BATEC. The WBATE estimator uses equal weights over the boundary grid points.}
\end{table}

The boundary-wide summaries reinforce the same interpretation. For the first stage, all 31 location-based pointwise estimates are statistically significant and all 31 uniform bands exclude zero; the distance-based first stage gives 30 pointwise significant estimates and 29 uniform bands excluding zero. Neither method finds statistically significant WBATE evidence for the reduced-form job-growth outcome or for the fuzzy effect. The location-based fuzzy WBATE is $-0.066$ with a $p$-value of $0.263$, while the distance-based fuzzy WBATE is $-0.099$ with a $p$-value of $0.124$.

The core empirical calls are short. The full script \code{CTY\_2026\_rd2d\_empapp.R} reads the clean dataset \code{CTY\_2026\_rd2d\_data.csv}, constructs the distance matrix, runs both analyses, and writes the figures and tables used in this section to the \code{inputs} folder.

\begin{lstlisting}[style=Rstyle]
fit_loc <- rd2d(Y = Y, X = X, assignment = T, b = b, fuzzy = W,
                fitmethod = "joint",
                params.cov = c("main", "itt", "fs"))
fit_dist <- rd2d.distance(Y = Y, distance = D, b = b, fuzzy = W,
                          fitmethod = "joint",
                          params.cov = c("main", "itt", "fs"))
summary(fit_loc, output = "main", cbands = "main")
summary(fit_dist, output = "main", cbands = "main")
\end{lstlisting}

\section{Conclusion}\label{sec: Conclusion}

This article introduced \pkg{rd2d}, software for causal inference in boundary discontinuity designs. The package implements location-based and distance-based local-polynomial methods, robust bias-corrected pointwise inference, uniform confidence bands, BATEC, WBATE, LBATE, fuzzy BATEC, fuzzy WBATE, and fuzzy LBATE. It also includes joint cluster-robust fitting, the earlier separate fitting convention, automatic bandwidth selection, and covariate efficiency adjustment. The current software ecosystem includes \proglang{R}, \proglang{Python}, and \proglang{Stata} implementations with aligned syntax and output conventions. The Opportunity Zones application illustrates how the package can separate the reduced-form BATEC, the first-stage BATEC, and the fuzzy BATEC along a multidimensional assignment boundary. Replication codes and related information are available at \url{https://rdpackages.github.io/}.

\section{Acknowledgments}

We thank Alberto Abadie, Kosuke Imai, Xinwei Ma and Filippo Palomba for insightful discussions. Cattaneo and Titiunik gratefully acknowledge financial support from the National Science Foundation (SES-2019432, DMS-2210561, SES-2241575 and SES-2342226). Cattaneo gratefully acknowledges financial support from the National Institute for Food and Agriculture (NIFA) through grant 2024-67023-42704, the Data-Driven Social Science initiative at Princeton University, and a 2026 Guggenheim Fellowship.

\appendix

\section{Multi-Platform Syntax}\label{sec: Multi Platform Syntax}

The three implementations share the same conceptual inputs. Users supply an outcome, a two-dimensional score or distance-to-boundary score, an assignment variable, evaluation points on the boundary, and optional fuzzy treatment receipt. In the code snippets below, \code{Z} denotes an optional matrix or data frame of predetermined covariates, and \code{cluster\_id} denotes an optional cluster identifier. The most visible differences across platforms are naming conventions and return objects.

\subsection{R}

In \proglang{R}, location-based methods use \code{rd2d()} and \code{rdbw2d()}, while distance-based methods use \code{rd2d.distance()} and \code{rdbw2d.distance()}.

\begin{lstlisting}[style=Rstyle]
install.packages("rd2d")
library(rd2d)

fit <- rd2d(Y = Y, X = X, assignment = T, b = b,
            fuzzy = W, params.cov = c("main", "itt", "fs"))
summary(fit, output = "main", cbands = "main")
summary(fit, output = "itt", cbands = "itt")
summary(fit, output = "fs", cbands = "fs")

fit_cov <- rd2d(Y = Y, X = X, assignment = T, b = b,
                fuzzy = W, cluster = cluster_id,
                covs.eff = Z, fitmethod = "joint",
                params.cov = c("main", "itt", "fs"))

fit_dist <- rd2d.distance(Y = Y, distance = D, b = b, fuzzy = W,
                          cluster = cluster_id,
                          covs.eff = Z, fitmethod = "joint",
                          params.cov = c("main", "itt", "fs"))
\end{lstlisting}

The \proglang{R} interface uses dotted option names such as \code{params.cov} and \code{covs.eff}, and returns S3 objects with \code{print()}, \code{summary()}, \code{plot()}, \code{coef()}, \code{vcov()}, and \code{confint()} methods. Uniform bands and aggregate summaries require the relevant covariance matrices to be stored through \code{params.cov} or related options. The main automatic bandwidth selectors are \code{mserd}, \code{cerrd}, \code{imserd}, \code{icerrd}, \code{msetwo}, \code{certwo}, \code{imsetwo}, and \code{icertwo}; user-supplied bandwidths are recorded as \code{"user provided"}. The default fitting convention is \code{fitmethod = "joint"}; \code{fitmethod = "separate"} requests the earlier side-specific convention. Distance-based kink handling uses \code{kink.unknown} and \code{kink.position}.

\subsection{Python}

The \proglang{Python} implementation mirrors the \proglang{R} API but follows Python naming conventions.

\begin{lstlisting}[style=Rstyle, language=Python]
from rd2d import rd2d, rdbw2d, rd2d_dist, rdbw2d_dist

fit = rd2d(Y, X, T, b, fuzzy=W,
           params_cov=["main", "itt", "fs"])
fit.summary(output="main", cbands="main")
fit.summary(output="itt", cbands="itt")
fit.summary(output="fs", cbands="fs")

fit_cov = rd2d(Y, X, T, b, fuzzy=W,
               cluster=cluster_id,
               covs_eff=Z, fitmethod="joint",
               params_cov=["main", "itt", "fs"])

fit_dist = rd2d_dist(Y, D, b=b, fuzzy=W,
                     cluster=cluster_id,
                     covs_eff=Z, fitmethod="joint",
                     params_cov=["main", "itt", "fs"])
\end{lstlisting}

The main differences are snake-case option names, such as \code{kernel_type}, \code{params_cov}, \code{covs_eff}, \code{kink_unknown}, and \code{kink_position}, and \proglang{Python} return objects that expose tabular outputs through familiar array or data-frame structures. The option \code{fitmethod="joint"} is the default, and \code{fitmethod="separate"} is available for comparisons with the earlier convention. Distance-based aliases use \code{rd2d_dist} and \code{rdbw2d_dist}; longer aliases are also available in the package.

\subsection{Stata}

The \proglang{Stata} implementation follows Stata command syntax. Location-based methods use \code{rd2d} and \code{rdbw2d}. Distance-based methods use \code{rd2d\_dist} and \code{rdbw2d\_dist}; aliases ending in \code{\_distance} are also available.

\begin{lstlisting}[style=Rstyle]
net install rd2d, from(https://raw.githubusercontent.com/rdpackages/rd2d/main/stata) replace

rd2d y x1 x2 assignment, b(0 0 0 1) fuzzy(w)
rdbw2d y x1 x2 assignment, b(0 0 0 1) fuzzy(w) bwparam(main)

rd2d y x1 x2 assignment, b(0 0 0 1) fuzzy(w) ///
    cluster(cluster_id) covseff(z1 z2) fitmethod(joint)

rd2d_dist y dist1 dist2, b(0 0 0 1) fuzzy(w) ///
    cluster(cluster_id) covseff(z1 z2) fitmethod(joint)
\end{lstlisting}

Stata stores results in \code{e()} matrices, including covariance matrices for the main, reduced-form, and first-stage tables when available. In fuzzy BD designs, the main table reports the fuzzy BATEC, while the reduced-form and first-stage BATECs are stored separately in \code{e(itt)} and \code{e(fs)}. The joint fitting convention is requested with \code{fitmethod(joint)}, the earlier side-specific convention with \code{fitmethod(separate)}, and covariate efficiency adjustment with \code{covseff()}. Distance-based kink options are written as \code{kinkunknown()} and \code{kinkposition()}.

To summarize the main syntax across platforms:
\begin{itemize}
    \item In \proglang{R}, the main commands are \code{rd2d()}, \code{rdbw2d()}, and \code{rd2d.distance()}. The interface returns S3 objects and uses dotted option names such as \code{params.cov} and \code{covs.eff}.
    \item In \proglang{Python}, the main commands are \code{rd2d()}, \code{rdbw2d()}, and \code{rd2d\_dist()}. The interface follows Python conventions with snake-case option names such as \code{params\_cov} and \code{covs\_eff}.
    \item In \proglang{Stata}, the main commands are \code{rd2d}, \code{rdbw2d}, and \code{rd2d\_dist}. Results are stored in \code{e()} returns, covariate adjustment is requested with \code{covseff()}, and distance-based aliases with \code{distance} are available.
\end{itemize}

\clearpage
\bibliographystyle{plainnat}
\bibliography{CTY_2026_rd2d--bib}

\end{document}